\documentclass[a4paper,11pt]{article}

\usepackage{jheppub} 

\usepackage{empheq,mathtools}
\usepackage{amsmath,amsfonts,amssymb}
\usepackage{mathrsfs}
\usepackage{cancel,slashed}
\usepackage{tensor}
\usepackage{xcolor}
\usepackage{graphicx}
\usepackage{subfigure}
\usepackage{float}

\makeatletter
\def\@fpheader{\relax}
\makeatother

\newcommand{\be}{\begin{equation}}
\newcommand{\ee}{\end{equation}}

\usepackage{ulem}

\usepackage[czech,english]{babel}
\usepackage{yfonts}

\usepackage{url}
\usepackage{mathabx}

\preprint{{PCFT-24-40, LITP-25-06}}

\usepackage{tikz}
\usetikzlibrary{positioning, arrows.meta}

\title{\boldmath Explorations of Universality in the Entropy and Hawking Radiation of Non-Extremal Kerr AdS$_4$ Black Holes}

\author[a, b]{Jun Nian,}
\author[c]{Leopoldo A. Pando Zayas,}
\author[a]{Wenni Zheng\,}

\affiliation[a]{International Centre for Theoretical Physics Asia-Pacific (ICTP-AP), University of Chinese Academy of Sciences (UCAS), 100190 Beijing, China}
\affiliation[b]{Peng Huanwu Center for Fundamental Theory, Hefei, Anhui 230026, China}
\affiliation[c]{Leinweber Institute for Theoretical Physics, University of Michigan, Ann Arbor, MI 48109, USA}

\emailAdd{nianjun@ucas.ac.cn}
\emailAdd{lpandoz@umich.edu}
\emailAdd{zhengwenni22@mails.ucas.ac.cn}

\abstract{ We comprehensively discuss various microscopic approaches to the Bekenstein-Hawking entropy for rotating, electrically charged, asymptotically AdS$_4$ {\it non-extremal} black holes in gauged supergravity. We apply the covariant phase space formalism to the near-horizon region to obtain a Cardy-formula-based microscopic explanation for the entropy, consistent with the Kerr/CFT correspondence. From the dual boundary CFT point of view, we estimate the free partition function in the matrix model approximation in the high-temperature regime and find qualitative agreement with the supergravity answer. All these different approaches match in the appropriate limits and support the universality of AdS black hole entropy even at high temperatures, far away from extremality. Prompted by the consistency of the results of statistical explanations for the AdS$_4$ black hole entropy, we discuss aspects of the rate of Hawking radiation at high temperatures from the CFT$_2$ perspective and found it to be universally proportional to the horizon area.}

\begin{document} 
\maketitle
\flushbottom

\section{Introduction}\label{sec:intro}

Understanding the microstates underlying the macroscopic black hole entropy is crucial in the quest to establish quantum gravity. The first microscopic explanation for black hole entropy was provided by considering specific asymptotically flat black holes~\cite{Strominger:1996sh, Callan:1996dv}, using elements of string theory and D-brane technology. For asymptotically AdS black holes, the  AdS/CFT correspondence~\cite {Maldacena:1997re, Gubser:1998bc, Witten:1998qj} provides a microscopic foundation for the Bekenstein-Hawking entropy via certain field-theoretic partition functions on the corresponding boundary conformal field theories (CFT). The first explicit computation of holographic microstate counting for higher-dimensional AdS black holes was performed by considering asymptotically AdS$_4$ supersymmetric magnetic and dyonic static black holes~\cite{Benini:2015eyy, Benini:2016rke} and the topologically twisted index~\cite{Benini:2015noa} of the dual ABJM theory~\cite{Aharony:2008ug} in the large-$N$ limit. There have also been substantial advances for supersymmetric, rotating, electrically charged AdS black holes. Microstate counting in the $\mathcal{N}=4$ SYM theory dual to AdS$_5$ BPS (supersymmetric and extremal) black hole were presented in~\cite{Hosseini:2017mds, Cabo-Bizet:2018ehj, Choi:2018hmj, Choi:2018vbz, Benini:2018ywd, Honda:2019cio, ArabiArdehali:2019tdm, Kim:2019yrz}. The counting in the three-dimensional ABJM theory provides a microscopic foundation for the entropy of AdS$_4$ black holes as described in~\cite{Choi:2018fdc, Choi:2019zpz, Nian:2019pxj, Benini:2019dyp, GonzalezLezcano:2022hcf, Bobev:2022wem, BenettiGenolini:2023rkq}; similar explanations are available in other dimensions~\cite{Hosseini:2018dob, Choi:2018fdc, Choi:2019miv, Nahmgoong:2019hko, Kantor:2019lfo, Crichigno:2020ouj}.

In the context of the original Strominger-Vafa computation \cite{Strominger:1996sh}, a simpler microscopic foundation can be obtained by considering the low-energy description of the near-horizon geometry \cite{Strominger:1997eq}. This formulation contains the seeds of a treatment that applies beyond the extremities. Indeed, in the near-horizon region, the Kerr/CFT correspondence~\cite{Guica:2008mu, Lu:2008jk, Hartman:2008pb, Chow:2008dp} allows us to reproduce the entropy of an extremal, rotating black hole utilizing a 2d CFT and its associated Cardy formula for microstate counting~\cite{Cardy:1986ie}. This formalism can be generalized to the near-extremal case~\cite{Bredberg:2009pv, Hartman:2009nz}. However, the non-extremal case is not as immediately explicit since it relies on properties of the phase space, such as the so-called hidden conformal symmetry~\cite{Castro:2010fd, Wang:2010qv}. More rigorously, the covariant phase space formalism should be applied to explain this symmetry~\cite{Haco:2018ske, Haco:2019ggi, Aggarwal:2019iay, Perry:2020ndy, Perry:2022udk} and obtain a microscopic description of the entropy of non-extremal black holes.

Motivated by the resounding success of microstate countings of extremal and near-extremal AdS black holes, we venture into aspects of universality in the far-from-extremality regime. A natural starting point is to consider the completely non-extremal Kerr black hole in AdS, whose temperature is arbitrary, especially the high-temperature limit. In this paper, we explore various approaches to the entropy of a specific non-extremal, rotating, electrically charged, asymptotically AdS$_4$ black hole. We extend previous approaches, known to work for extremal zero-temperature and near-extremal small-temperature black holes, to the case of arbitrary temperature. From the near-horizon point of view, we apply the covariant phase space formalism to the derivation of black hole entropy and find that our result matches the Bekenstein-Hawking entropy. On the boundary side, we compute the free partition function of 3d ABJM theory under a Cardy-like limit. Although the result in this case cannot precisely reproduce the black hole entropy, it provides the correct scaling in $N$ and temperature. These two microscopic descriptions, either from the AdS/CFT correspondence on the boundary or from the Kerr/CFT correspondence in the near-horizon region, can reproduce the Bekenstein-Hawking entropy for rotating AdS black holes both in the BPS case~\cite{David:2020ems} and the near-extremal case~\cite{Nian:2020qsk, David:2020jhp}. Together with the original computation in gravity, these three approaches describe the black hole entropy from different perspectives, as shown in Fig.~\ref{graph}.

\begin{figure}[ht]
  \centering

\begin{tikzpicture}[
    node distance=2.5cm,
    box/.style={rectangle, draw, fill=orange!30, rounded corners, minimum width=6cm, minimum height=1.5cm, align=center},
    arrow/.style={-Stealth}
]

\node[box] (ads) {\large  $\text{AdS}_{d+1} \text{ black hole} \ (S^{(1)}_{\text{BH}})$};
\node[box, right=of ads] (boundary) {\large $\text{boundary CFT}_{d} \ (S^{(3)}_{\text{BH}})$};
\node[box, below=1.5cm of ads] (warped) {\large$\text{Warped AdS}_{3} \text{ geometry}$};
\node[box, right=of warped] (cft2){\large$\text{CFT}_{2} \ (S^{(2)}_{\text{BH}})$};

\draw[arrow] (ads) -- node[above, font=\large] {\large \text{AdS/CFT}} (boundary);
\draw[arrow] (warped)-- node[above]  {\large\text{Kerr/CFT}}(cft2);
\draw[arrow] (ads) --(warped);

\end{tikzpicture}
\caption{\label{graph} Three approaches to the universal result of asymptotically AdS rotating black holes}
\end{figure}

The universality of fluid dynamics lies in its ability to describe the large wavelengths, long timescales, and high-temperature behavior of most systems. From the point of view of the fluid/gravity correspondence, fluid mechanics is also applicable to describe the non-extremal black holes we focus on in this paper. Following \cite{Bhattacharyya:2007vs}, we review the hydrodynamic interpretation of the non-extremal AdS$_4$ black hole entropy to facilitate comparison with direct field theory computations. Finally, having explored the universality results of non-extremal AdS$_4$ black hole entropy, we provide a microscopic derivation of the Hawking radiation rate in the near-horizon region of these black holes, in the same spirit as \cite{Callan:1996dv, Nian:2020qsk, David:2020jhp, Nian:2023xmr}.

This paper is organized as follows. We review some basic properties of electrically charged rotating AdS$_4$ black holes and the Kerr/CFT correspondence in Sec.~\ref{sec:black hole solution} and Sec.~\ref{sec:Kerr/CFT}, respectively, emphasizing essential features relevant to our discussion in this paper. In Sec.~\ref{sec:soft hair}, we apply the covariant phase space formalism to reproduce the Bekenstein-Hawking entropy. To discuss the non-extremal AdS$_4$ black hole entropy from the boundary theory, we first review the dual hydrodynamic description in Sec.~\ref{sec:fluid/gravity}. Then, in Sec.~\ref{sec:boundary}, we discuss the black hole entropy from the free partition function of ABJM theory on the boundary. Having seen certain scaling universality of non-extremal AdS$_4$ black holes from different approaches, we discuss the microscopic computation of near-horizon Hawking radiation rate in Sec.~\ref{sec:Hawking radiation}. Finally, we conclude with some discussions and open questions in Sec.~\ref{sec:discussion}.

\section{Rotating electrically charged AdS$_4$ black hole background}\label{sec:black hole solution}

In this section, we introduce the asymptotically AdS$_4$ Kerr-Newman black hole with rotation, two sets of pairwise electric charges under the $U(1)\times U(1)$ subgroup of SO(4)-gauged $\mathcal{N}=4$ supergravity~\cite{Chong:2004na}. The solution to the metric $g_{\mu\nu}$, the gauge fields $A_1$, $A_2$ and the scalar fields $\varphi_1$, $\chi_1$ equations of motions is given by
\begin{align}
  ds^2 & = - \frac{\Delta_r}{W} \left(dt - \frac{a\, \textrm{sin}^2 \theta}{\Xi} d\phi \right)^2 + W \left(\frac{dr^2}{\Delta_r} + \frac{d\theta^2}{\Delta_\theta} \right) + \frac{\Delta_\theta\, \textrm{sin}^2 \theta}{W} \left[a\, dt - \frac{r_1 r_2 + a^2}{\Xi} d\phi \right]^2\, ,\label{eq:AdS4Metric}\\
    A_1 & = \frac{2 \sqrt{2} m\, \textrm{sinh} (\delta_1)\, \textrm{cosh} (\delta_1)\, r_2}{W} \left(dt - \frac{a\, \textrm{sin}^2 \theta}{\Xi} d\phi \right) + \alpha_{41}\, dt\, ,\\
    A_2 & = \frac{2 \sqrt{2} m\, \textrm{sinh} (\delta_2)\, \textrm{cosh} (\delta_2)\, r_1}{W} \left(dt - \frac{a\, \textrm{sin}^2 \theta}{\Xi} d\phi \right) + \alpha_{42}\, dt\, ,\\
    e^{\varphi_1} & = \frac{r_1^2 + a^2\, \cos^2 \theta}{W}=1+\frac{r_1(r_1-r_2)}{W}\, ,\\
    \chi_1 & = \frac{a (r_2 - r_1)\, \textrm{cos}\, \theta}{r_1^2 + a^2\, \textrm{cos}^2 \theta}\, ,
  \end{align}
  where
  \begin{align}
  \begin{split}
    r_i & \equiv r + 2 m\, \textrm{sinh}^2 (\delta_i)\, ,\quad (i = 1, 2)\\
    \Delta_r & \equiv r^2 + a^2 - 2 m r + g^2 r_1 r_2 (r_1 r_2 + a^2)\, ,\\
    \Delta_\theta & \equiv 1 - g^2 a^2\, \textrm{cos}^2 \theta\, ,\\
    W & \equiv r_1 r_2 + a^2\, \textrm{cos}^2 \theta\, ,\\
    \Xi & \equiv 1 - a^2 g^2\, ,
  \end{split}
  \end{align}
  and $g \equiv \ell_4^{-1}$ denotes the inverse of the AdS$_4$ radius. The parameter space for this black hole is $(a,m,\delta_1,\delta_2)$. The temperature, angular velocity, and electrostatic potentials measured in the asymptotically rotating frame are given by~\cite{Cvetic:2005zi}
  \begin{align}
    T_{\text{H}}=\frac{\Delta_r'(r_+)}{4\pi\bigl(r_1(r_+)\, r_2(r_+)+a^2\bigr)}\,, & \quad  \Omega_{\text{H}} = \frac{a\bigl(1 + g^2\, r_1(r_+)\, r_2(r_+)\bigr)}{r_1(r_+)\, r_2(r_+) + a^2}\,,\\
    \Phi_1=\Phi_2=\frac{2m\sinh(\delta_1) \cosh(\delta_1)\, r_2(r_+)}{r_1(r_+)\, r_2(r_+) + a^2}\,, & \quad \Phi_3 = \Phi_4 = \frac{2m\sinh(\delta_2) \cosh(\delta_2)\, r_1(r_+)}{r_1(r_+)\, r_2(r_+)+a^2}\,.
  \end{align}
  The Bekenstein-Hawking entropy is given by 
  \begin{align}
    S_{\text{BH}}=\frac{\pi\,\hat{r}^2_+}{\Xi}\,,\label{Bekenstein-Hawking entropy}
  \end{align}
  where
  \begin{align}
    \hat{r}_\pm^2 \equiv r_1(r_\pm)\, r_2(r_\pm)+a^2\, ,
  \end{align}
  with $r_\pm$ denoting the radii of the outer and inner horizons determined by the positive roots of the blackening factor $\Delta_r(r_\pm)=0$. The corresponding charges, namely, the energy, the angular momentum, and the electric charges, are
   \begin{align}
    E=\frac{m}{\Xi^2}\Bigl(1+\sinh^2(\delta_1)+\sinh^2(\delta_2)\Bigr)\,, & \quad J=\frac{m\,a}{\Xi^2}\Bigl(1+\sinh^2(\delta_1)+\sinh^2(\delta_2)\Bigr)\,,\\
    Q_1=Q_2=\frac{m\sinh(\delta_1)\cosh(\delta_1)}{2\Xi}\,, & \quad Q_3=Q_4=\frac{m\sinh(\delta_2)\cosh(\delta_2)}{2\Xi}\,.
   \end{align}

  To describe thermodynamics, it is more convenient to replace $m$ with the outer horizon $r_+$. If we consider the large black hole defined by taking the large horizon radius limit, i.e., the horizon radius $r_+$ of the black hole is large compared to the AdS radius $\ell_4$, with $k_i = 2m\sinh^2 r_i/r_+ (i=1,2)$ fixed, the parameter $m$ can be expressed as 
  \begin{align}
    m=\frac{g^2r_+^3}{2}(1+k_1)^2(1+k_2)^2\,.\label{eq: m}
  \end{align}
  The thermodynamic quantities of these AdS$_4$ Kerr-Newman black holes can be simplified as 
  \begin{align}
   & T_{\text{H}}=\frac{g^2r_+}{4\pi}\bigl(3 + k_1 + k_2  - k_1 k_2\bigr)\,,\quad \Omega_{\text{H}}=a\,g^2\,,\label{eq:TH in k from gravity side}\\
   & \Phi_1=\Phi_2=g\,r_+\sqrt{k_1}(1+k_2)\,,  \quad \Phi_3=\Phi_4=g\,r_+\sqrt{k_2}(1+k_1)\,,\label{eq:phi in k from gravity side}\\
   & E=\frac{g^2r_+^3}{2\Xi^2}(1+k_1)^2(1+k_2)^2\,,  \quad J=\frac{g^2r_+^3a}{2\Xi^2}(1+k_1)^2(1+k_2)^2\,,\label{eq:AdS4 BH energy}\\
   & Q_1=Q_2=\frac{gr_+^2(1+k_1)(1+k_2)\sqrt{k_1}}{4\Xi}\,,  \quad Q_3=Q_4=\frac{gr_+^2(1+k_1)(1+k_2)\sqrt{k_2}}{4\Xi}\,,\label{eq:Q}\\
   & S_{\text{BH}} =  \, \frac{\pi r_+^2(1+k_1)(1+k_2)}{\Xi}\,.\label{Bekenstein-Hawking entropy2}
  \end{align}

Later in this paper, we consider a special class of AdS$_4$ black holes at high temperature, under the assumption that the dual boundary description is well approximated by contributions only from supersymmetric operators restricted to the singlet sector via a matrix integration. 

The supersymmetric (SUSY) constraint for the black hole is enforced by the constraint ~\cite{Chong:2004na} 
\begin{align}
  e^{2\delta_1+2\delta_2}=1+\frac{2}{ag}\,,\label{eq:SUSY condition}
\end{align}
while the zero-temperature condition is defined by $\Delta'_r(r_+)=0$, which takes the form 
\begin{align}
    mg=\frac{\cosh(\delta_1+\delta_2)}{e^{(\delta_1+\delta_2)/2}\sinh^2(\delta_1+\delta_2)\sinh(2\delta_1)\sinh(2\delta_2)}\,.\label{eq: zero-temperature condition}
\end{align}
The BPS condition corresponds to the intersection of the SUSY constraint~\eqref{eq:SUSY condition} and the zero-temperature condition~\eqref{eq: zero-temperature condition}, as illustrated in Fig.~\ref{fig:parameter space}.

\begin{figure}
    \begin{center}
    \begin{tikzpicture}[x={(2cm,0.5cm)}, y={(1.cm,-1.cm)}, z={(0cm,2cm)}]
    \fill[blue!10] (0,0,0) -- (4,0,0) -- (4,2,0) -- (0,2,0) -- cycle;
    \fill[red!10] (2,0,-1.6) -- (2,0,2) -- (2,2,2) -- (2,2,-1.6) -- cycle;
    \draw[blue, thick] (0,0,0) -- (4,0,0) -- (4,2,0) -- (0,2,0) -- cycle;
    \draw[red, thick] (2,0,-1.6) -- (2,0,2) -- (2,2,2) -- (2,2,-1.6) -- cycle;
  
    \draw[thick, black, line width=2pt] (2,0,0) -- (2,2,0);

    \node[font=\bfseries\color{blue}] at (3.5,1.65,0)  {$\mathbf{T=0}$};
      \node[font=\bfseries\color{red}] at (2,0.7,1.5) {SUSY};
       \node[font=\bfseries\color{black}] at (1.75,1,0) {BPS};
\end{tikzpicture}
  \end{center}
  \caption{The codimension-1 supersymmetric hypersurface (red plane) and the codimension-1 zero-temperature hypersurface (blue plane) in the parameter space; their intersection defines the BPS sector (black solid line).}\label{fig:parameter space}
  \end{figure}

\section{Kerr/CFT correspondence}\label{sec:Kerr/CFT}

This section briefly reviews the Kerr/CFT correspondence relevant to the next section. The Kerr black hole~\cite{Kerr:1963ud} in four-dimensional asymptotically flat spacetime can be written in Boyer-Lindquist coordinates as
\begin{align}
  ds^2=-\frac{\Delta}{\rho^2}\bigl(dt-a\,\sin^2\theta d\phi\bigr)^2+\frac{\sin^2\theta}{\rho^2}\Bigl((r^2+a^2)d\phi-adt\Bigr)^2+\frac{\rho^2}{\Delta}dr^2+\rho^2d\theta^2\,,
\end{align}
\begin{align}
  \Delta\equiv r^2-2Mr+a^2\,,\qquad \rho^2\equiv r^2+a^2\cos^2\theta\,.
\end{align}
We have taken $c=\hbar=G_N=1$. The Kerr metric above is parameterized by the mass $M$ and the angular momentum $J=M\,a$. The horizons are located at 
\begin{align}
  r_\pm=M\pm\sqrt{M^2-a^2}\,.
\end{align}
The Hawking temperature and the angular velocity of the horizon are respectively given by 
\begin{align}
  T_{\text{H}}=\frac{r_+-M}{4\pi Mr_+}\,,\qquad \Omega_\text{H}=\frac{a}{2Mr_+}\,.
\end{align}
In order to study the near-horizon region of the extremal Kerr geometry, we consider the Bardeen-Horowitz scaling~\cite{Bardeen:1999px}
\begin{align}
  \hat{t}=\frac{\lambda t}{2M}\,,\quad x=\frac{r-M}{\lambda M}\,,\quad \hat{\phi}=\phi-\frac{t}{2M}\,,
\end{align}
with $\lambda\to0$, and find the near-horizon extreme Kerr (NHEK) in Poincar\'e-type coordinates as
\begin{align}
  ds^2=2J\Omega^2\left(-x^2d\hat{t}^2+\frac{dx^2}{x^2}+d\theta^2+\Lambda^2(d\hat{\phi}+xd\hat{t})^2\right)\,,
\end{align}
where
\begin{align}
    \Omega^2=\frac{1+\cos^2\theta}{2}\,,\quad \Lambda=\frac{2\sin\theta}{1+\cos^2\theta}\,.
\end{align}
After imposing appropriate boundary conditions, the most general diffeomorphisms that preserve the asymptotic form of the metric have the generators:
\begin{align}
  \zeta_\epsilon=\epsilon_n(\phi)\partial_\phi-r\epsilon'_n(\phi)\partial_r\,,\quad \text{with}\quad \epsilon_n(\phi)=-e^{-in\phi}\,.
\end{align}
These diffeomorphisms obey the SL$(2,\mathbb{R})$ Lie algebra, whose quantum version is the Virasoro algebra. The Dirac bracket of conserved charges $Q_\zeta$  generated by the above generators gives rise to a central charge $c_{\text{L}}=12J$. 
With the left and the right temperatures of the extremal Kerr geometry
\begin{align}
  T_{\text{L}}=\frac{1}{2\pi}\,,\quad T_{\text{R}}=0\,,
\end{align}
the Bekenstein-Hawking entropy is reproduced microscopically via the Cardy formula 
\begin{align}
  S=\frac{\pi^2}{3}c_{\text{L}}T_{\text{L}}=2\pi J\, .
\end{align}
Besides the asymptotically flat Kerr black holes, the Kerr/CFT correspondence has been generalized to asymptotically AdS extremal Kerr-Newman black holes \cite{Lu:2008jk, David:2020ems}. Moreover, the extremal Kerr/CFT can be generalized to the near-extremal case by introducing the near-NHEK geometry \cite{Bredberg:2009pv, Hartman:2009nz, Nian:2020qsk, David:2020jhp}.

The SL$(2,\mathbb{R})$ algebra that is manifest in the near-horizon geometry is not explicitly realized in the full black hole geometry. It can be revealed, however, through the properties of the Klein-Gordon equation for a massless scalar
\begin{align}
  \frac{1}{\sqrt{-g}}\partial_\mu\Bigl(\sqrt{-g}\,g^{\mu\nu}\partial_\nu\Phi(t,r,\theta,\phi)\Bigr)=0\,.
\end{align}
Uncovering an SL$(2,\mathbb{R})$ algebra within the above equation has been dubbed ``hidden conformal symmetry''~\cite{Castro:2010fd}. Due to the translation invariance in both the time $t$ and the azimuthal angle $\phi$, the scalar field $\Phi$ can be naturally expanded as
\begin{align}
  \Phi(t,r,\theta,\phi)=e^{-i\omega t+im\phi}R(r)S(\theta)\,.
\end{align}
Thus, the Klein-Gordon equation is separated into two second-order linear ordinary differential equations, the angular equation and the radial equation: 
\begin{align}
  \biggl[\frac{1}{\sin\theta}\partial_\theta(\sin\theta\,\partial_\theta)-\frac{m^2}{\sin^2\theta}+\omega^2a^2\cos^2\theta\biggr]S(\theta)=-K_\ell S(\theta)\,,\label{the angular equation}
\end{align}
\begin{align}
  \biggl[\partial_r\Delta\partial_r+\frac{(2Mr_+\omega-m\,a)^2}{(r-r_+)(r_+-r_-)}-\frac{(2Mr_-\omega-m\,a)^2}{(r-r_-)(r_+-r_-)}+\bigl(r^2+2M(r+2M)\bigr)\omega^2\biggr]R(r)=K_\ell R(r)\,,\label{the radial equation}
\end{align}
where $K_\ell$ denotes a real separation constant. For an arbitrary frequency $\omega$, the angular equation~\eqref{the angular equation} cannot be solved exactly, and $K_\ell$ is a function of $(\ell,\, m,\, a\,\omega)$ that can only be determined numerically.

At low frequencies, where the scalar excitation is much smaller than the inverse of the curvature radius of the Kerr black hole, the equation \eqref{the angular equation} is approximated by the confluent Heun equation with $K_\ell=\ell(\ell+1)+\mathcal{O}(a^2\omega^2),\  m\in (-\ell, \ell)$.\footnote{ Recent developments (see e.g. \cite{Aminov:2020yma, Bonelli:2021uvf}) allow an exact treatment of the wave equation beyond the low-frequency regime.} In the near region $r\ll 1/\omega$, the last term  proportional to $\omega^2$ in the angular equation~\eqref{the angular equation} can be neglected, and \eqref{the angular equation} reduces to the standard ordinary spherical harmonic equation on $S^2$:
\begin{align}
  \nabla^2_{S^2}S(\theta)=\biggl[\frac{1}{\sin\theta}\partial_\theta(\sin\theta\partial_\theta)-\frac{m^2}{\sin^2\theta}\biggr]S(\theta)=-\ell(\ell+1)S(\theta)\,,\quad m=-\ell,\cdots,\ell\,.
\end{align}
We are interested in the radial equation~\eqref{the radial equation}, which is also a Heun equation in general. In the near region, the last term proportional to $\omega^2$ in this equation can also be omitted, and thus, \eqref{the radial equation} becomes a hypergeometric equation:
\begin{align}
  \biggl[\partial_r\Delta\partial_r + \frac{(2Mr_+\omega-m\,a)^2}{(r-r_+)(r_+-r_-)} - \frac{(2Mr_-\omega-m\,a)^2}{(r-r_-)(r_+-r_-)}\biggr] R(r) = \ell(\ell+1) R(r)\,.
\end{align}
The solution can be expressed as 
\begin{align}
  R(r) = & \, C_1\left(\frac{r-r_+}{r-r_-}\right)^{-\frac{i(\omega-m\Omega_{\text{H}})}{4\pi T_{\text{H}}}}(r-r_-)^{-1-\ell} \nonumber\\
  &\times\,_{2}F_1\biggl[1+\ell-i\frac{4M}{r_+-r_-}(M\omega-r_+m\Omega_{\text{H}}),1+\ell-i2M\omega;1-\frac{i(\omega-m\Omega_{\text{H}})}{2\pi T_{\text{H}}};\frac{r-r_+}{r-r_-}\biggr]\nonumber\\
  &+C_2\left(\frac{r-r_+}{r-r_-}\right)^{\frac{i(\omega-m\Omega_{\text{H}})}{4\pi T_{\text{H}}}}(r-r_-)^{-1-\ell} \nonumber\\
  &\times\,{_2}F_1\biggl[1+\ell+i\frac{4M}{r_+-r_-}(M\omega-r_+m\Omega_{\text{H}}),1+\ell+i2M\omega;1+\frac{i(\omega-m\Omega_{\text{H}})}{2\pi T_{\text{H}}};\frac{r-r_+}{r-r_-}\biggr]\,,
\end{align}
where $C_1$ and $C_2$ are two constants determined by the boundary conditions. According to the relation between Lie groups and differential equations, hypergeometric functions transform in representations of SL$(2,\mathbb{R})$. This implies that the conformal symmetry is encoded in the near region of the radial sector of the Klein-Gordon wave equation for a massless scalar in the Kerr black hole background.

One way to make the underlying SL$(2,\mathbb{R})$ symmetry more manifest uses the following convenient coordinates: 
\begin{align}
  &\omega^+=\sqrt{\frac{r-r_+}{r-r_-}}e^{2\pi T_{\text{R}}\phi}\,,\\
  &\omega^-=\sqrt{\frac{r-r_+}{r-r_-}}e^{2\pi T_{\text{L}}\phi-\frac{t}{2M}}\,,\\
  &y=\sqrt{\frac{r_+-r_-}{r-r_-}}e^{\pi(T_{\text{L}}+T_{\text{R}})\phi-\frac{t}{4M}}\,,
\end{align}
with 
\begin{align}
  T_{\text{L}}\equiv\frac{r_++r_-}{4\pi a}\,,\quad T_{\text{R}}\equiv\frac{r_+-r_-}{4\pi a}\,,
\end{align}
two copies of local vector fields 
\begin{align}
\begin{split}
  H_1&=i\partial_+\,,\\
  H_0&=i\left(\omega^+\partial_++\frac{1}{2}y\partial_y\right),\\
  H_{-1}&=i\Bigl((\omega^+)^2\partial_++\omega^+y\partial_y-y^2\partial_-\Bigr)\,,
\end{split}
\end{align}
and 
\begin{align}
\begin{split}
  \bar{H}_1&=i\partial_-\,,\\
  \bar{H}_0&=i\left(\omega^-\partial_-+\frac{1}{2}y\partial_y\right),\\
  \bar{H}_{-1}&=i\Bigl((\omega^-)^2\partial_-+\omega^-y\partial_y-y^2\partial_+\Bigr)\,,
\end{split}
\end{align}
obey the SL$(2,\mathbb{R})_{\text{L}}\times$SL$(2,\mathbb{R})_{\text{R}}$ Lie algebra
\begin{align}
  &[H_0, H_{\pm 1}]=\mp iH_{\pm 1}\,,\quad [H_{-1},H_1]=-2iH_0\,,\nonumber\\
  &[\bar{H}_0, \bar{H}_{\pm 1}]=\mp i\bar{H}_{\pm 1}\,,\quad [\bar{H}_{-1},\bar{H}_1]=-2i\bar{H}_0\,.
\end{align} 
With this symmetry, we can again obtain the Kerr black hole entropy using the Cardy formula in two-dimensional conformal field theory 
\begin{align}
  S=\frac{\pi^2}{3}\bigl(c_{\text{L}}T_{\text{L}}+c_{\text{R}}T_{\text{R}}\bigr)=2\pi Mr_+\,,
\end{align}
which matches the Bekenstein-Hawking entropy result from the gravity side.

It is worth noting that recently, certain radial and angular equations, similar to those presented above, have been mapped to the confluent Heun equation, leading to exact solutions of various perturbation problems, including greybody factors and quasinormal modes \cite{Aminov:2020yma, Bonelli:2021uvf}. It would be interesting to leverage these connections with other problems in mathematical physics, not only to obtain exact solutions but also to provide a deeper conceptual understanding of the Kerr/CFT correspondence and its underlying SL$(2,\mathbb{R})$ symmetry.

\section{Covariant Phase Space Formalism}\label{sec:soft hair}

The hidden conformal symmetry of a non-extremal Kerr black hole, reviewed in Sec.~\ref{sec:Kerr/CFT}, can be more rigorously studied using the covariant phase space formalism \cite{Haco:2018ske, Haco:2019ggi, Aggarwal:2019iay, Perry:2020ndy, Perry:2022udk}. The authors of \cite{Haco:2018ske} utilized the phase space formalism to interpret the central charges of the 2d conformal field theory dual to the Kerr black hole as surface integrals. This approach is based on the fact that the diffeomorphisms Vir$_\text{L}\otimes$Vir$_\text{R}$ act non-trivially on the horizon of the black hole and generate soft hair. This formalism was extended to the Kerr-Newman black hole in \cite{Haco:2019ggi} and further adapted to incorporate a negative cosmological constant to accommodate asymptotically AdS$_4$ black holes in \cite{Perry:2020ndy}. Other backgrounds, including Kerr-Newman-AdS and Kerr-Taub-NUT black holes, were treated in \cite{Perry:2022udk}. 
Besides the Einstein gravity, the covariant phase space formalism can also be applied to higher-curvature gravity theories (see e.g. \cite{Ghodrati:2016vvf}).

In this section, we reproduce the Bekenstein-Hawking entropy for the non-extremal AdS$_4$ Kerr-Newman black holes, including an additional dilaton and an axion field introduced in Sec.~\ref{sec:black hole solution} microscopically via this formalism. After introducing the conformal coordinates, a set of diffeomorphisms acting non-trivially on the horizon of the black hole forms a pair of centreless Virasoro algebras. The central terms of these Virasoro algebras are encoded in the covariant charges of the generators, from which the central charges can be extracted. This allows us to compute the black hole entropy using the Cardy formula for the degeneracy of states in a CFT$_2$.

\vspace{0.5cm}
  \noindent{\bf Conformal coordinates}
\vspace{0.3cm}

  \noindent The conformal coordinates $(\omega^\pm, y)$ are
  \begin{align}
    \begin{split}
      &\omega^+=R(r)\,e^{2\pi T_{\text{R}}\phi}\,,\\
      &\omega^-=R(r)\,e^{2\pi T_{\text{L}}\phi-\Delta_r'(r_+)t/r_*^2}\,,\\
      &y=Q(r)\,e^{\pi(T_{\text{L}}+T_{\text{R}})\phi-\Delta_r'(r_+)t/(2r_*^2)}\,,
    \end{split}
  \end{align}
  where $R(r)^2=(r-r_+)/\Delta_r'(r_+)$, and $Q(r)$ is chosen to satisfy $R^2(r)+Q^2(r)=1$.
  The left and the right temperatures are defined as 
  \begin{align}
    T_{\text{L}}\equiv \frac{\Delta_r'(r_+)}{4\pi\,a\,\Xi}\frac{\hat{r}^2_++\hat{r}_-^2}{r_*^2}\,,\qquad T_{\text{R}}\equiv \frac{\Delta_r'(r_+)}{4\pi\,a\,\Xi}\,,\label{left-right temperature}
  \end{align}
  where 
  \begin{align}
    r_*^2 \equiv \hat{r}^2_+-\hat{r}^2_-\,.
  \end{align}
  The bifurcation surface is located at $\omega^\pm (r) = 0$. The corresponding inverse transformation is 
  \begin{align}
    \begin{split}
      &t=\frac{1}{\Delta_r'(r_+)}\biggl(\hat{r}_+^2\ln\frac{\omega^+}{\omega^-}+\hat{r}_-^2\ln\bigl(\omega^+\omega^-+y^2\bigr)\biggr),\\
      &r=r_++\frac{\omega^+\omega^-\Delta_r'(r_+)}{\omega^+\omega^-+y^2}\,,\\
      &\phi=\frac{a\Xi}{\Delta_r'(r_+)}\ln\frac{\omega^+(\omega^+\omega^-+y^2)}{\omega^-}\,.
    \end{split}
  \end{align}
  Under this transformation, the metric of the leading and the subleading orders around the bifurcation surface becomes
  \begin{align}
    \begin{split}
      ds^2=&\frac{4W_+}{y^2}d\omega^+d\omega^-+\frac{4\Delta_\theta\,r_*^4\, a^2\sin^2\theta}{\Delta_r'(r_+)^2\,y^2\,W_+} dy^2+\frac{W_+}{\Delta_\theta}d\theta^2\\
      &-\frac{4\omega^+ }{y^3}\biggl(r_*^2-\frac{\Delta_\theta\, a^2\sin^2\theta}{W_+\Delta_r'(r_+)^2}\bigl(r_*^4-\Delta_r'(r_+)\,r_*^2\,\tilde{r}_+\bigr)\biggr)d\omega^-dy\\
      &-\frac{4\omega^-}{y^3}\biggl(W_++W_--\frac{\Delta_\theta\, a^2\sin^2\theta}{W_+\Delta_r'(r_+)^2}\Bigl(r_*^4+\Delta'_r(r_+)\,r_*^2\,\tilde{r}_+\Bigr)\biggr)d\omega^+dy+\cdots\,,
    \end{split}
  \end{align}
  where the ellipsis denotes the corrections at least second order in $\omega^\pm$, and 
  \begin{align}
    W_\pm & = r_1(r_\pm)r_2(r_\pm)+a^2\cos^2\theta\,,\\
  \tilde{r}_+ & = r_1(r_+)+r_2(r_+)\,.
\end{align}
The volume element is 
\begin{align}
  \epsilon_{\theta+-y}=\frac{4W_+r_*^2\,a\,\sin\theta}{\Delta_r'(r_+)\,y^3}+\cdots\,.
\end{align}
For later convenience, we find the relation
\begin{align}
  \Gamma^+_{+y}+\Gamma^-_{-y}=-\frac{2}{y}\,.
\end{align}

\vspace{0.5cm}
\noindent{\bf Conformal vector fields}
\vspace{0.3cm}

\noindent We consider the conformal vector fields
\begin{equation}
  \zeta(\varepsilon)=\varepsilon\,\partial_++\frac{1}{2}(\partial_+\varepsilon)y\,\partial_y\,,\qquad \bar{\zeta}(\bar{\varepsilon})=\bar{\varepsilon}\,\partial_-+\frac{1}{2}(\partial_-\bar{\varepsilon})y\,\partial_y\,,
\end{equation}
where $\varepsilon\equiv\varepsilon(\omega^+)$ and $\bar{\varepsilon}\equiv\bar{\varepsilon}(\omega^-)$. The fact that the vector fields $\zeta$ and $\bar{\zeta}$ should be invariant under $2\pi$ azimuthal rotations restricts $\varepsilon$ and $\bar{\varepsilon}$ to be 
\begin{equation}
  \varepsilon_n=2\pi T_{\text{R}}(\omega^+)^{1+\frac{in}{2\pi T_{\text{R}}}}\,,\qquad \bar{\varepsilon}_n=2\pi T_{\text{L}}(\omega^-)^{1+\frac{in}{2\pi T_{\text{L}}}}\,.
\end{equation}
Then, the corresponding vector fields, $\zeta_n\equiv\zeta(\varepsilon_n)$ and $\bar{\zeta}_n\equiv\bar{\zeta}(\bar{\varepsilon}_n)$, are considered as a left-right pair of centreless Virasoro algebras, i.e.,
\begin{align}
  [\zeta_m,\, \zeta_n]=i(n-m)\zeta_{n+m}\,,\qquad [\bar{\zeta}_m,\,\bar{\zeta}_n]=i(n-m)\bar{\zeta}_{n+m}\,.
\end{align}

\vspace{0.5cm}
\noindent{\bf Covariant charges}
\vspace{0.3cm}

\noindent The conformal coordinates and the diffeomorphisms we have established above give rise to the covariant linearized charges $\delta\mathcal{Q}_n=\delta\mathcal{Q}(\zeta_n,\mathcal{L}_{\zeta_m};g)$. The central charge term $K_{m,n}$ in the Virasoro charge algebra for the right movers\footnote{The left mover associated with $\bar{\zeta}_n$ follows the same scheme.}
\begin{align}
  \{\mathcal{Q}_n,\mathcal{Q}_m\}=(m-n)\mathcal{Q}_{m+n}+K_{m,n}
\end{align}
is described by the linearized charge $\delta\mathcal{Q}$ associated with the vector fields $\zeta_n$ acting on the horizon
\begin{align}
  K_{m,n}=\delta \mathcal{Q}(\zeta_n,\mathcal{L}_{\zeta_m}g;g)=\frac{c_{\text{R}}\,m^3}{12}\delta_{m+n}\,.
\end{align}
In general, the linearized charge contains both the Iyer-Wald part and the Wald-Zoupas boundary counterterm\footnote{The Wald-Zoupas counterterm is required for the integrability and associativity of the charge algebra.}
\begin{align}
  \delta\mathcal{Q}=\delta\mathcal{Q}_{\text{IW}}+\delta\mathcal{Q}_{\text{ct}}\,.\label{linear covariant charge}
\end{align}
These linearized charges can be expressed as surface integrals on the horizon using the phase space formalism. The Iyer-Wald charge is given by
\begin{align}
  \delta\mathcal{Q}_{\text{IW}}=\frac{1}{16\pi}\int_{\partial\Sigma}*F_{\text{IW}}\,,
\end{align}
where
\begin{align}
  F_{\text{IW}}^{ab}=\frac{1}{2}\nabla^a\zeta^b\,h+\nabla^a h^{cb}\zeta_c+\nabla_c\zeta^ah^{cb}+\nabla_ch^{ca}\zeta^b-\nabla^ah\,\zeta^b-(a\,\leftrightarrow\,b)\,.
\end{align}
Here, the variation $h_{ab}$ is defined by $g_{ab}\to g_{ab}+h_{ab}$, where the metric perturbation $(h_m)_{ab}=\mathcal{L}_{\zeta_m}g_{ab}$ and $h=h_{ab}\,g^{ab}$. It turns out that the leading term that contributes to the Iyer-Wald term on the bifurcation surface is 
\begin{align}
  F_{\text{IW}}^{-y}=4\,h^{y-}_m\,\zeta^y_n\,\Gamma^-_{-y}\,.\label{IW-y}
\end{align}
For later convenience, we find that 
\begin{align}
  h_m^{-y}=g^{+-}\partial_+\zeta_m^y=\frac{y^3\varepsilon_m''}{4W_+}\,,\qquad \text{with} \ \ \varepsilon_m''\equiv\partial_+^2\varepsilon\,.
\end{align}

For the counterterm 
\begin{align}
  \delta\mathcal{Q}_{\text{ct}}=\frac{1}{16\pi}\int_{\partial\Sigma}F_{\text{ct}}^{ab}d\Sigma_{ab}
\end{align}
with
\begin{align}
  F_{\text{ct}}^{ab}=-2\tensor{N}{_d^c}\nabla_c\bigl(\zeta^ah^{db}\bigr) - (a\,\leftrightarrow\,b)\,,
\end{align}
we consider 
\begin{align}
  \tensor{N}{_+^+}=1\,,\qquad \tensor{N}{_-^-}=-1\, ,
\end{align}
and find the only non-trivial term that contributes to the Wald-Zoupas charge is 
\begin{align}
  F_{\text{ct}}^{-y}=2\,h_m^{-y}\zeta_n^y\bigl(\Gamma^+_{+y}-\Gamma^-_{-y}\bigr)\,.\label{ct-y}
\end{align}
Combining the terms from the Iyer-Wald part and the Wald-Zoupas counterterm results in
\begin{align}
 \begin{split}
  \delta\mathcal{Q}&=\frac{1}{16\pi}\int d\theta d\omega^+\epsilon_{\theta+-y}\bigl(F_{\text{IW}}^{-y}+F_{\text{ct}}^{-y}\bigr)\\
  &=i\frac{ar_*^2}{2\Delta_r'(r_+)}m^3\delta_{m+n,0}\,.
 \end{split}
\end{align}
From the Dirac bracket to the commutator $\{.\,,.\}_{\text{DB}}\to-i[.\,,.]$, we can read off the right-moving central charge 
\begin{align}
  c_{\text{R}}=\frac{6\,a\,r_*^2}{\Delta_r'(r_+)}=\frac{6a}{\Delta_r'(r_+)}\bigl(r_1(r_+)r_2(r_+)-r_1(r_-)r_2(r_-)\bigr)\,.\label{central charge}
\end{align}
The computation of the left-moving sector is similar to the above procedure and gives the same result of $c_{\text{L}}$ as $c_{\text{R}}$.

\vspace{0.5cm}
\noindent{\bf Area law}
\vspace{0.3cm}

\noindent Now, we can apply the Cardy formula for degeneracy of states in CFT$_2$~\cite{Cardy:1986ie} to compute the non-extremal AdS$_4$ Kerr-Newman black hole entropy:
\begin{align}
  S_{\text{Cardy}} = \frac{\pi^2}{3}\bigl(c_{\text{L}}\,T_{\text{L}}+c_{\text{R}}\,T_{\text{R}}\bigr)\, ,
\end{align}
where the left and the right temperatures can be obtained using \eqref{left-right temperature}:
\begin{align}
  \hspace*{-0.7cm} T_{\text{L}} & = \frac{\Delta_r'(r_+)}{4\pi a\Xi}\frac{\hat{r}_+^2+\hat{r}_-^2}{r_*^2}\nonumber\\
  \hspace*{-0.7cm} & = \frac{\Delta_r'(r_+)\Bigl[\bigl(r_++2m\sinh^2(\delta_1)\bigr)\bigl(r_++2\sinh^2(\delta_2)\bigr)+\bigl(r_-+2m\sinh^2(\delta_1)\bigr)\bigl(r_-+2m\sinh^2(\delta_2)\bigr)+2a^2\Bigr]}{4\pi a(1-a^2g^2)\Bigl[\bigl(r_++2m\sinh^2(\delta_1)\bigr)\bigl(r_++2m\sinh^2(\delta_2)\bigr)-\bigl(r_-+2m\sinh^2(\delta_1)\bigr)\bigl(r_-+2m\sinh^2(\delta_2)\bigr)\Bigr]}\,, \label{eq:TL}\\
  \hspace*{-0.7cm} T_{\text{R}} & =\frac{\Delta_r'(r_+)}{4\pi a\Xi}=\frac{\Delta_r'(r_+)}{4\pi a(1-a^2g^2)}\,, \label{eq:TR}
\end{align}
while the central charges $c_{L, R}$ are given by \eqref{central charge}:
\begin{align}
  {} & c_{\text{L}}=c_{\text{R}}\nonumber\\
  =\, & \frac{6\,a\,r_*^2}{\Delta_r'(r_+)}=\frac{6a}{\Delta_r'(r_+)}\bigl(r_1(r_+)r_2(r_+)-r_1(r_-)r_2(r_-)\bigr)\nonumber\\
  =\, & \frac{6a}{\Delta_r'(r_+)}\Bigl(\bigl(r_++2m\sinh^2(\delta_1)\bigr)\bigl(r_++2m\sinh^2(\delta_2)\bigr)-\bigl(r_-+2m\sinh^2(\delta_1)\bigr)\bigl(r_-+2m\sinh^2(\delta_2)\bigr)\Bigr)\,. \label{eq:cLcR}
\end{align}
Hence, the black hole entropy is 
\begin{align}
  S_{\text{Cardy}}=\frac{\pi\,\hat{r}^2_+}{\Xi}=\frac{\pi\Bigl[\bigl(r_1+2m\sinh^2(\delta_1)\bigr)\bigl(r_++2m\sinh^2(\delta_2)\bigr)+a^2\Bigr]}{1-a^2g^2}\,,
\end{align}
which is precisely the same as the Bekenstein-Hawking entropy~\eqref{Bekenstein-Hawking entropy} from the gravity side. Note that the entropy in~\eqref{Bekenstein-Hawking entropy} need not be near extremality.

\section{Entropy from Fluid/Gravity Duality}\label{sec:fluid/gravity}

In what follows we review the black hole entropy in the large $r_+$ limit from the boundary conformal fluid mechanics along the lines of~\cite{Bhattacharyya:2007vs}, which states that the thermodynamics of large AdS$_4$ black holes is equivalent to stationary solutions of the relativistic Navier-Stokes equation of conformal fluid in local thermodynamic equilibrium on $\mathbb{R}\times \text{S}^2$.

In fluid dynamics, the conservation of the stress tensor and the charge currents, $\nabla_\mu T^{\mu\nu}=0$ and $\nabla_\mu J_i^\mu=0$, play an essential role.
To proceed, one implements a derivative expansion of the stress tensor, the charge current, and the entropy current. The zeroth order (perfect fluid) of these quantities is described in terms of the local proper energy density $\rho$, local charge densities $r_i$, and fluid velocity $u^\mu=\gamma(1,\vec{v})$ governed by Lorentz invariance and the appropriate static limit
\begin{align}
  &T^{\mu\nu}_0=\rho u^\mu u^\nu+\mathcal{P}P^{\mu\nu}\,,\\
  &J_{i,0}^{\mu}=r_iu^\mu\,,\\
  &J_{S,0}^\mu=s u^\mu\,,
\end{align}
where $P^{\mu\nu}=g^{\mu\nu}+u^\mu u^\nu$ is the projection tensor which projects vectors onto S$^2$ in the fluid rest frame. At the subleading order, dissipation and diffusion appear. The stress tensor and the currents in the first subleading order are
\begin{align}
  &T^{\mu\nu}_1=-\zeta\vartheta P^{\mu\nu}-2\eta\sigma^{\mu\nu}+q^\mu u^\nu+u^\mu q^\nu\,,\\
  &J^\mu_{i,1}=q_i^\mu\,,\\
  &J^\mu_{S,1}=\frac{1}{\mathcal{T}}(q^\mu-\mu_iq_i^\mu)\,,
\end{align}
where $\zeta$ is the bulk viscosity,\footnote{The bulk viscosity $\zeta$ of the conformal fluid vanishes.} $\eta$ is the shear viscosity, $\mathcal{T}$ is the local temperature, and $\mu_i$ are the chemical potentials. The expansion parameter, shear tensor, heat flux, and diffusion current are given by
\begin{align}
  &\vartheta=\nabla_\mu u^\mu\,,\\
  &\sigma^{\mu\nu}=\frac{1}{2}(P^{\mu\lambda}\nabla_\lambda u^\nu+P^{\nu\lambda}\nabla_\lambda u^\mu)-\frac{1}{2}\vartheta P^{\mu\nu}\,,\\
  &q^\mu=-\kappa P^{\mu\nu}(\partial_\nu\mathcal{T}+a_\nu \mathcal{T})\,,\\
  &q_i^\mu=-D_{ij}P^{\mu\nu}\partial_\nu\left(\frac{\mu_j}{\mathcal{T}}\right)\,,
\end{align}
with the thermal conductivity $\kappa$, the diffusion coefficients $D_{ij}$, and the acceleration $a^\mu=u^\nu\nabla_\nu u^\mu$. For a stationary fluid, $\sigma^{\mu\nu}$, $q^\mu$ and $q{_i^\mu}$ must vanish. In equilibrium, the fluid motion should be just a rigid rotation due to $\sigma^{\mu\nu}=0$, which gives $u^\mu=\gamma(1,0,\omega_1)$ with $\gamma=(1-v^2)^{-1/2}$ and $v^2=\omega_1^2\sin^2\theta$, while $q^\mu=0$ and $q_i^\mu=0$ lead to $\mathcal{T}=\tau\gamma$ and $\mu_i=\mathcal{T}\nu_i$.

For conformal fluid systems, the thermodynamic potential is defined as 
\begin{align}
  \Phi\equiv\mathcal{E}-\mathcal{T}\mathcal{S}-\mu_i\mathcal{R}_i\,,\quad i=1,\cdots,4\,,
\end{align}
where $\mathcal{E}$ and $\mathcal{S}$ are the rest energy and the entropy of the conformal fluid, respectively, and we denote by $\mathcal{R}_i$  the fluid $\mathcal{R}$-charges with $\mu_i$ being the corresponding chemical potentials. The first law of thermodynamics of this system is given by
\begin{align}
  d\Phi=-\mathcal{S}d\mathcal{T}-\mathcal{P}dV-\mathcal{R}_id\mu_i\, .
\end{align}
Conformal invariance and extensivity require the thermodynamic potential to be
\begin{align}
  \Phi=-\text{vol}(\text{S}^2)\mathcal{T}^dh(\nu_i)\,,\label{thermodynamic potential}
\end{align}
where $h$, as a function of the ratio between the  chemical potentials $\mu_i$ and the local temperature $\mathcal{T}$, is defined precisely by the relation \eqref{thermodynamic potential}. The thermodynamic quantities, including the rest energy density, the pressure,\footnote{The relation between the energy density and the pressure is determined by conformal invariance, which requires the stress tensor to be traceless.} and the entropy density, can be expressed with the function $h(\mu_i)$ as follows
\begin{align}
  &\rho=2\mathcal{P}=2\mathcal{T}^3h(\nu)\,,\\
  &r_i=\mathcal{T}^2h_i(\nu)\,,\\
  &s=\mathcal{T}^2\,(3h-\nu_ih_i)\,,
\end{align}
where $h_i=(\partial h)/(\partial \nu_i)$. The conserved charges are then integrated to be
\begin{align}
  &E=\frac{2\text{vol}(\text{S}^2)\tau^3h(\nu)}{(1-\omega_1^2)^2}\,,\\
  &L_1=\frac{2\text{vol}(\text{S}^2)\tau^3h(\nu)\omega_1}{(1-\omega_1^2)^2}\,,\\
  &R_i=\frac{\text{vol}(\text{S}^2)\tau^2h_i(\nu)}{1-\omega_1^2}\,,\\
  &S=\frac{\text{vol}(\text{S}^2)\tau^2[3h(\nu)-\nu_ih_i(\nu)]}{1-\omega_1^2}\,.
\end{align}
The corresponding chemical potentials are obtained following standard thermodynamic relations:
\begin{align}
  &T=\left(\frac{\partial E}{\partial S}\right)_{L_1,R_i}=\tau\,,\\
  &\Omega_1=\left(\frac{\partial E}{\partial L_1}\right)_{S,R_i}=\omega_1\,,\\
  &\zeta_i=\left(\frac{\partial E}{\partial R_i}\right)_{S,L_1,R_j}=\tau\nu_i\,.
\end{align}
The grand-canonical partition function is thus given by
\begin{align}\label{eq:def fluid partition function}
  \mathcal{Z}=&\exp\biggl[-\frac{E-TS-\Omega_1L_1-\zeta_iR_i}{T}\biggr]=\exp\biggl[\frac{\text{vol}(\text{S}^2)T^2h(\zeta/T)}{1-\Omega_1^2}\biggr]\,.
\end{align}

For the AdS$_4$ black hole we are considering here in the large horizon limit $r_+\gg\ell_4$, the functions $h(\nu_i)$ and $h_i$ in the dual conformal fluid on $\mathbb{R}\times$S$^2$ are given by~\cite{Harmark:1999xt, Bhattacharyya:2007vs}
\begin{align}
  h & = \frac{\mathcal{P}}{\mathcal{T}^3} = \frac{4\pi^2(1+k_1)^2(1+k_2)^2}{\Xi\, g^4(3+k_1+k_2-k_1k_2)^3}\,,\\
  h_1 & \equiv\frac{\partial h}{\partial\nu_1}=\frac{r_1}{\mathcal{T}^2} = \frac{\pi(1+k_1)(1+k_2)\sqrt{k_1}}{\Xi\, g^3(3+k_1+k_2-k_1k_2)^2}\,,\quad h_2\equiv\frac{\partial h}{\partial\nu_2}=\frac{r_2}{\mathcal{T}^2}=h_1\,,\\
  h_3 & \equiv\frac{\partial h}{\partial\nu_3}=\frac{r_3}{\mathcal{T}^2}=\frac{\pi(1+k_1)(1+k_2)\sqrt{k_2}}{\Xi\, g^3(3+k_1+k_2-k_1k_2)^2}\,, \quad h_4\equiv\frac{\partial h}{\partial\nu_4}=\frac{r_4}{\mathcal{T}^2}=h_3\,,\\
  \nu_1 & = \nu_2 = \frac{4 \pi (1 + k_2) \sqrt{k_1}}{3 + k_1 + k_2 - k_1 k_2}\, ,\quad \nu_3 = \nu_4 = \frac{4 \pi (1 + k_1) \sqrt{k_2}}{3 + k_1 + k_2 - k_1 k_2}\, .\label{eq:expression of nu_i}
\end{align}
All the thermodynamics can thus be obtained following the conformal fluid language. In particular, the Bekenstein-Hawking entropy in the large-$r_+$ limit is
\begin{align}
  \mathcal{S}=\text{vol}(S^2)\cdot s=\frac{\pi r_+^2(1+k_1)(1+k_2)}{\Xi}\,,
\end{align}
which is the same as the result from the gravity side \eqref{Bekenstein-Hawking entropy2}, but now endowed with a description of degrees of freedom in the dual fluid dynamics. The grand-canonical partition function can be expressed in terms of the thermodynamic quantities as 
\begin{align}
  \log \mathcal{Z}=\frac{16\pi^3(2N)^{3/2} T^2 (1+k_1)^2 (1+k_2)^2}{3 (1-\Omega_1^2) (3 + k_1 + k_2 - k_1 k_2)^3}\, .\label{eq: grand-canonical partition function}
\end{align}

In summary, there is a universality in the rotating AdS black hole free energy obtained at high temperature, which is compatible with the fluid/gravity correspondence. We expect to reproduce this universal free energy from the field theory perspective by explicitly calculating the partition function. In the next section, we will turn directly to the partition function of the boundary field theory and make a comparison with the result \eqref{eq: grand-canonical partition function}.

\section{Boundary Field Theory Partition Function}\label{sec:boundary}

We now explore black hole entropy from the perspective of boundary field theory. Recall that the AdS$_4$ black hole solution we have discussed in previous sections is dual to the three-dimensional $\mathcal{N}=6$ ABJM theory with gauge group U($N$)$_k\times$ U($N$)$_{-k}$ on S$^1\times$S$^2$~\cite{Aharony:2008ug}. The superconformal symmetry algebra of this theory is Osp$(6|4)$, which has the bosonic subalgebra SO(3,2)$\times$ SO(6). The SO(3,2)$\,\simeq\, $Sp$(4)$ is the three-dimensional conformal symmetry, while the $R$-symmetry is SO(6)$\,\simeq\,$SU(4), whose Cartan generators are denoted by $h_{1, 2, 3}$. In addition, there is a baryonic symmetry denoted as U(1)$_b$. This theory contains two gauge fields $(A_\mu, \tilde{A}_\mu)$ and matter fields, including four complex scalars $Y^A$ as well as four Weyl fermions $\psi_A$, with $A=1,\cdots,4$, in the bifundamental representation of the gauge group. For details about the global charges of each field, see, for example, \cite{Bhattacharya:2008bja, Kim:2009wb}. The corresponding partition function with supercharges $Q$ and $S$ is defined in terms of a weighted trace over the Hilbert space $\mathcal{H}$ of the states on $S^2$ in radial quantization as   (note that we do not insert $(-1)^F$, where $F$ is the fermion number,  in the trace)
\begin{align}
  Z(q,q';y_1,y_2)={\text{Tr}}_{\mathcal{H}}\Bigl[e^{-\beta'\{Q,S\}} e^{-\beta(\epsilon+j_3)} e^{-\gamma_1h_1-\gamma_2h_2}\Bigr],\label{eq: def of partition function}
\end{align}
where $\beta'$, $\beta$, and $\gamma_{1, 2}$ stand for the chemical potentials, whose corresponding fugacities are defined as 
\begin{align}
  q\equiv e^{-\beta}\,,\quad q'\equiv e^{-\beta'}\,,\quad  y_1\equiv e^{-\gamma_1}\,,\quad y_2\equiv e^{-\gamma_2}\,.
\end{align}
According to the algebra of supercharges in this theory, we have $e^{-\beta'\{Q, S\}}e^{-\beta(\epsilon + j_3)} = e^{-(\beta' + \beta)\epsilon + \beta'h_3 - (\beta - \beta') j_3}$. Thus, we can define the inverse physical temperature in this theory as $\bar{\beta}=\beta+\beta'$, and view $\beta-\beta'$ as a chemical potential associated with $j_3$, denoted as $\bar{\omega}\equiv\beta-\beta'$. Later on in this paper, we will use $\lambda_i$ to denote other chemical potentials in general.

 In evaluating the partition function~\eqref{eq: def of partition function}, a full accounting of the non-BPS spectrum in the interacting ABJM theory is presently intractable. We therefore focus on the BPS sector, which is protected by supersymmetry and allows for a controlled microstate count via an index-like partition function, with the $(-1)^F$ factor removed. Our approach is motivated by a specific physical question: does the scaling of entropy obtained from this protected sector survive in the high-temperature, far-from-extremality regime? We find that the BPS-derived scaling 
$S\sim N^{3/2}T^2$ matches the gravitational result, which suggests that the universal macroscopic behavior is encoded already in the kinematic, representation-theoretic data of the theory. This provides a non-trivial consistency check of universality, rather than a complete dynamical derivation from the full Hilbert space.

\subsection{The Matrix Model}

Evaluating the partition function~\eqref{eq: def of partition function} is a rather formidable task since it is not protected by supersymmetry, since we have not inserted $(-1)^F$ in the trace. In principle, the above partition function depends on the couplings of the theory. It might, as in the case of QCD, capture different phases at different values of the thermodynamic potentials. In this section, we merely estimate the above partition function by considering only certain protected operators and further imposing the singlet condition with respect to the full gauge group. This is a major simplification, but has been shown to capture interesting non-trivial physics in previous examples \cite{Aharony:2003sx, Nishioka:2008gz}.  In particular, \cite{Nishioka:2008gz} has considered the $\mathcal{N}=6$ ABJM theory on $S^1 \times S^2$ in the limit $N \to \infty$ and $k \to \infty$, while in this paper we focus on the limit $N \to \infty$ at finite $k$.

The matrix model that we seek to evaluate for the ABJM partition function can be obtained from the general form of the supersymmetric one~\cite{Kim:2009wb, Choi:2019zpz, GonzalezLezcano:2022hcf}:
\begin{align}
\label{Eq:PartitionFunction}
  Z(q,q';y_1,y_2) & = \frac{1}{(N!)^2}\sum_{\mathfrak{m},\tilde{\mathfrak{m}}\in\mathbb{Z}^N}\int\biggl[\prod_{i=1}^N\frac{d\alpha_i}{2\pi}\frac{d\tilde{\alpha}_i}{2\pi}\biggr]\prod_{\substack{i<j\\\mathfrak{m}_i=\mathfrak{m}_j}}\biggl[2\sin\Bigl(\frac{\alpha_i-\alpha_j}{2}\Bigr)\biggr]^2\prod_{\substack{i<j\\\tilde{\mathfrak{m}}_i=\tilde{\mathfrak{m}}_j}}\biggl[2\sin\Bigl(\frac{\tilde{\alpha}_i-\tilde{\alpha}_j}{2}\Bigr)\biggr]^2\nonumber\\
   {} & \qquad\qquad\qquad\qquad \times Z_0\, Z_{\text{cl}}\,Z_{\text{ch}}\,Z_{\text{g}}\,,
  \end{align}
 where the classical term, $Z_{\rm cl}$, the 1-loop part from the chiral multiplets, $Z_{\rm ch}$, the 1-loop part from the vector multiplets, $Z_{\rm g}$, and the Casimir energy, $Z_0$, in the partition function are given respectively by
 \begin{align}
  Z_{\text{cl}} & = e^{ik\sum_{i=1}^N(\mathfrak{m}_i\alpha_i-\tilde{\mathfrak{m}}_i\tilde{\alpha}_i)}\, ,\\
  Z_{\text{ch}} & = \prod_{i,j=1}^N\exp\biggl[\sum_{n=1}^\infty\frac{1}{n}\Bigl(f^+_{ij}(q^n,q'^n;y_1^n,y_2^n)e^{in(\tilde{\alpha}_j-\alpha_i)}+f^-_{ij}(q^n,q'^n;y_1^n,y_2^n)e^{in(\alpha_i-\tilde{\alpha}_j)}\Bigr)\biggr]\, ,\\
  Z_{\text{g}} & = \prod_{i,j=1}^N\exp\biggl[\sum_{n=1}^\infty\frac{1}{n}\Bigl(f_{ij}^{\text{adj}}(q^n,q'^n)e^{-in(\alpha_i-\alpha_j)}+\tilde{f}_{ij}^{\text{adj}}(q^n,q'^n)e^{-in(\tilde{\alpha}_i-\tilde{\alpha}_j)}\Bigr)\biggr]\, ,\\
  Z_0 & = \prod_{i,j=1}^N\exp\Bigl[-\beta\Bigl(|\mathfrak{m}_i-\tilde{\mathfrak{m}}_j|-\sum_{i<j}|\mathfrak{m}_{i}-\mathfrak{m}_{j}|-\sum_{i<j}|\tilde{\mathfrak{m}}_{i}-\tilde{\mathfrak{m}}_{j}|\Bigr)\Bigr]\, .
 \end{align}
In the case of a free theory, it is efficient to construct the partition function starting from the single-trace, single-particle or single-letter partition functions, denoted generically by $f_{ij}$ above. Then, we plethystically  exponentiate them to obtain the full partition function of the theory. We will proceed with this approach, even though the interaction cannot be neglected for the partition function. In equation \eqref{Eq:PartitionFunction}, to account for the various group representations that appear in the multi-particle words, one further projects to the singlet sector by integrating over the Haar measure of the gauge group. The single-particle partition functions over the $ij$-component for each pair of scalars and fermions in representations $(N,\tilde{N})$ and $(\tilde{N}, N)$ are written as\footnote{Note that $\jmath$ denotes the angular momentum quantum numbers, while $i$ and $j$ are the matrix indices associated with the gauge group. Here, we turn off the fugacity for monopole operators for simplicity.}
\begin{align}
  f_{ij}^{\pm {\text{B}}}(q,q';y_1,y_2) & \equiv\sum_{\substack{4\, {\text{scalars}}\\s=\pm 1}}\sum_{\jmath=\frac{|\mathfrak{m}_i-\tilde{\mathfrak{m}}_j|}{2}}^\infty\sum_{j_3=-\jmath}^\jmath\left(q^{\epsilon_{\jmath}+j_3}q'^{\epsilon_{\jmath}-h_3-j_3}y_1^{h_1}y_2^{h_2}\right)\, ,\\
  f_{ij}^{\pm {\text{F}}}(q,q';y_1,y_2) & \equiv\sum_{4\,{\text{fermions}}}\sum_{\jmath=\frac{|\mathfrak{m}_i-\tilde{\mathfrak{m}}_j|\pm 1}{2}}^\infty\sum_{j_3=-\jmath}^\jmath\left(q^{\epsilon_{\jmath}+j_3}q'^{\epsilon_{\jmath}-h_3-j_3}y_1^{h_1}y_2^{h_2}\right)\, .
\end{align}
The results turn out to be explicitly as follows \cite{Kim:2009wb}
\begin{align}
  f_{ij}^{+{\text{B}}}(q,q';y_1,y_2) & = \left(\sqrt{\frac{y_1}{y_2}}+\sqrt{\frac{y_2}{y_1}}\right)\sum_{\jmath=\frac{|\mathfrak{m}_i-\tilde{\mathfrak{m}}_j|}{2}}^\infty q^{\frac{1}{2}}\,\sum_{l=0}^{2\jmath}q'^{2\jmath-l}q^{l} \nonumber\\
  {} & \quad + \left(\sqrt{y_1y_2}+\frac{1}{\sqrt{y_1y_2}}\right)\sum_{\jmath=\frac{|\mathfrak{m}_i-\tilde{\mathfrak{m}}_j|}{2}}^\infty q'q^{\frac{1}{2}}\,\sum_{l=0}^{2\jmath}q'^{2\jmath-l}q^{l}\, ,\\
  f_{ij}^{-{\text{B}}}(q,q';y_1,y_2) & = \left(\sqrt{y_1y_2}+\frac{1}{\sqrt{y_1y_2}}\right)\sum_{\jmath=\frac{|\tilde{\mathfrak{m}}_i-\mathfrak{m}_j|}{2}}^\infty q^{\frac{1}{2}}\sum_{l=0}^{2\jmath}q'^{2\jmath-l}q^l \nonumber\\
  {} & \quad + \left(\sqrt{\frac{y_1}{y_2}}+\sqrt{\frac{y_2}{y_1}}\right)\sum_{\jmath=\frac{|\tilde{\mathfrak{m}}_i-\mathfrak{m}_j|}{2}}^\infty q'q^{\frac{1}{2}}\sum_{l=0}^{2\jmath}q'^{2\jmath-l}q^l \, ,\\
  f_{ij}^{+{\text{F}}}(q,q';y_1,y_2) & = -\left(\sqrt{y_1y_2}+\frac{1}{\sqrt{y_1y_2}}\right)\sum_{\jmath=\frac{|\mathfrak{m}_i-\tilde{\mathfrak{m}}_j|+1}{2}}^\infty q^{\frac{1}{2}}\sum_{l=0}^{2\jmath}q'^{2\jmath-l}q^{l} \nonumber\\
  {} & \quad - \left(\sqrt{\frac{y_1}{y_2}}+\sqrt{\frac{y_2}{y_1}}\right)\sum_{\jmath=\frac{|\mathfrak{m}_i-\tilde{\mathfrak{m}}_j|-1}{2}}^\infty q'q^{\frac{1}{2}}\sum_{l=0}^{2\jmath}q'^{2\jmath-l}q^l\, ,\\
  f_{ij}^{-{\text{F}}}(q,q';y_1,y_2) & = -\left(\sqrt{\frac{y_1}{y_2}}+\sqrt{\frac{y_2}{y_1}}\right)\sum_{\jmath=\frac{|\mathfrak{m}_i-\tilde{\mathfrak{m}}_j|+1}{2}}q^{\frac{1}{2}}\sum_{l=0}^{2\jmath}q'^{2\jmath-l}q^{l} \nonumber\\
  {} & \quad - \left(\sqrt{y_1y_2}+\frac{1}{\sqrt{y_1y_2}}\right)\sum_{\jmath=\frac{|\mathfrak{m}_i-\tilde{\mathfrak{m}}_j|-1}{2}}^\infty q'q^{\frac{1}{2}}\sum_{l=0}^{2\jmath}q'^{2\jmath-l}q^{l}\, .
\end{align}
The single letters for the vector multiplet read 
\begin{align}
  f_{ij}^{\text{adj,B}}(q,q') & = \sum_{\jmath=\frac{|\mathfrak{m}_{ij}|}{2}+1}\frac{q^{2\jmath+1}-q'^{2\jmath+1}}{q-q'}+\sum_{\jmath=\frac{|\mathfrak{m}_{ij}|}{2}}\frac{qq'\bigl(q^{2\jmath+1}-q'^{2\jmath+1}\bigr)}{q-q'}+(1-\delta_{\mathfrak{m}_{ij}})\frac{q'q^{|\mathfrak{m}_{ij}|}-qq'^{|\mathfrak{m}_{ij}|}}{q-q'}\, ,\\
  f_{ij}^{\text{adj,F}}(q,q') & = \sum_{\jmath=\frac{|\mathfrak{m}_{ij}|+1}{2}}\frac{q'\bigl(q^{2\jmath+1}-q'^{2\jmath+1}\bigr)}{q-q'}-\sum_{\jmath=\frac{|\mathfrak{m}_{ij}|+1}{2}}q\frac{q^{2\jmath+1}q'^{2\jmath+1}}{q-q'}-(1-\delta_{\mathfrak{m}_{ij}})\frac{q\bigl(q^{\mathfrak{m}_{ij}}-q'^{\mathfrak{m}_{ij}}\bigr)}{q-q'}\, ,
\end{align}
and similar expressions for $\tilde{f}_{ij}^{\text{adj,B}}(q,q')$ and $\tilde{f}_{ij}^{\text{adj,F}}(q,q')$.

 Using the above single-letter partition functions and under the simplifications we have stated, the full partition function can be obtained by summing over all multi-trace states, i.e., the Fock space of all single-trace states. Note that compared to the ABJM theory's superconformal index (SCI) \cite{Choi:2019zpz, GonzalezLezcano:2022hcf, Kim:2009wb}, the relative sign between the bosonic and the fermionic contributions has been removed due to the absence of the factor $(-1)^F$ in the definition of the partition function. This is the object we will evaluate at high temperature and compare with the results of black hole thermodynamics.

Next, we replace the fugacities $y_1$ and $y_2$ with $t_i\, (i=1,\cdots,4)$ as in \cite{Choi:2019zpz} through the following relations:
\begin{align}
      t_1=\frac{1}{\sqrt{y_1y_2}}\,,\quad t_2=\sqrt{y_1y_2}\,,\quad t_3=\sqrt{\frac{y_2}{y_1}}\,,\quad t_4=\sqrt{\frac{y_1}{y_2}}\,.
     \end{align}
Consequently, we have the building blocks for the chiral and the vector multiplets
 \begin{align}
  {} & f_{ij}^{+}(q^n,q'^n;t_i^n) \nonumber\\
  =\, & f_{ij}^{+{\text{B}}}(q^n,q'^n;t_i^n)+(-1)^{n-1}f_{ij}^{+{\text{F}}}(q^n,q'^n;t_i^n)\nonumber\\
  =\, & \frac{t_{3,4}^n}{q^n-q'^n}\left(\frac{q^{n(1/2+|\mathfrak{m}_i-\tilde{\mathfrak{m}}_j|)}\bigl(q^n+(-1)^nq'^n\bigr)}{1-q^{2n}}-\frac{q^{n/2}q'^{n(1+|\mathfrak{m}_i-\tilde{\mathfrak{m}}_j|)}\bigl(1+(-1)^n\bigr)}{1-q'^{2n}}\right)\nonumber\\
  {} & + \frac{t_{1,2}^{n}}{q^n-q'^n}\left(\frac{q^{n(3/2+|\mathfrak{m}_i-\tilde{\mathfrak{m}}_j|)}(q'^n+(-1)^nq^n)}{1-q^{2n}}-\frac{q^{n/2}q'^{n(2+|\mathfrak{m}_i-\tilde{\mathfrak{m}}_j|)}(1+(-1)^n)}{1-q'^{2n}}\right)\, ,\\
  {} & {} \nonumber\\
  {} & f_{ij}^{-}(q^n,q'^n;t_i^n) \nonumber\\
  =\, & f_{ij}^{-{\text{B}}}(q^n,q'^n;t_i^n)+(-1)^{n-1}f_{ij}^{-{\text{F}}}(q^n,q'^n;t_i^n)\nonumber\\
  =\, & \frac{t_{1,2}^{-n}}{q^n-q'^n}\left(\frac{q^{n(1/2+|\mathfrak{m}_i-\tilde{\mathfrak{m}}_j|)}(q^n+(-1)^nq'^n)}{1-q^{2n}}-\frac{q^{n/2}q'^{n(1+|\mathfrak{m}_i-\tilde{\mathfrak{m}}_j|)}(1+(-1)^n)}{1-q'^{2n}}\right)\nonumber\\
  {} & + \frac{t_{3,4}^{-n}}{q^n-q'^n}\left(\frac{q^{n(3/2+|\mathfrak{m}_i-\tilde{\mathfrak{m}}_j|)}(q'^n+(-1)^nq^n)}{1-q^{2n}}-\frac{q^{n/2}q'^{n(2+|\mathfrak{m}_i-\tilde{\mathfrak{m}}_j|)}(1+(-1)^n)}{1-q'^{2n}}\right)\, ,
\end{align}
\begin{align}
{} & f_{ij}^{\text{adj}}(q^n,q'^n)= f_{ij}^{\text{adj,B}}(q^n,q'^n)+(-1)^{n-1}f_{ij}^{\text{adj,F}}(q^n,q'^n) \nonumber\\
  =\, & \left(\frac{q^{n(|\mathfrak{m}_{ij}|+2)}}{1-q^{2n}}-\frac{q'^{n(|\mathfrak{m}_{ij}|+2)}}{1-q'^{2n}}\right)\frac{q^n+q'^n}{q^n-q'^n}\bigl(1+(-1)^n\bigr) \nonumber\\
  {} & +\frac{1-\delta_{\mathfrak{m}_{ij}}}{q^n-q'^n}\Bigl(q^{n|\mathfrak{m}_{ij}|}\bigl(q'^n+(-1)^nq^n\bigr)-q^nq'^{n|\mathfrak{m}_{ij}|}\bigl(1+(-1)^n\bigr)\Bigr)\, ,\\
  {} & {} \nonumber\\
{} & \tilde{f}_{ij}^{\text{adj}}(q^n,q'^n)= \tilde{f}_{ij}^{\text{adj,B}}(q^n,q'^n)+(-1)^{n-1}\tilde{f}_{ij}^{\text{adj,F}}(q^n,q'^n)\nonumber\\
  =\, & \Bigl(\frac{q^{n(|\tilde{\mathfrak{m}}_{ij}|+2)}}{1-q^{2n}}-\frac{q'^{n(|\tilde{\mathfrak{m}}_{ij}|+2)}}{1-q'^{2n}}\Bigr)\frac{q^n+q'^n}{q^n-q'^n}\bigl(1+(-1)^n\bigr)\nonumber\\
  {} & + \frac{1-\delta_{\tilde{\mathfrak{m}}_{ij}}}{q^n-q'^n}\Bigl(q^{n|\tilde{\mathfrak{m}}_{ij}|}\bigl(q'^n+(-1)^nq^n\bigr)-q^nq'^{n|\tilde{\mathfrak{m}}_{ij}|}\bigl(1+(-1)^n\bigr)\Bigr)\, .
\end{align}
Since we are interested in the large-$N$ limit eventually, in which $i$ and $j$ will be identified, the second term in $f_{ij}^{\text{adj}}(q^n,q'^n)$ and $\tilde{f}_{ij}^{\text{adj}}(q^n,q'^n)$ will be neglected hereafter.
Differently from the SCI~\cite{Kim:2009wb}, $q'$ in the free partition function cannot be eliminated; instead, $q=e^{-\beta}$ and $q'=e^{-\beta'}$ are both considered as fugacities. In the SCI case, by contrast,  $q'$ ultimately does not appear in the result.

Let us highlight the conceptual difference between the SCI and the free partition function in more detail. The SCI can be defined from the UV description as a partition function on $S^1\times S^2$ up to the Casimir energy factor. From the definition of SCI, the standard insertion of $(-1)^F$  in the trace results in cancellations for each boson-fermion pair except for the BPS states. Thus, only the BPS states contribute to the trace, and the index is independent of the inverse temperature and the coupling constants. In contrast, the partition function contains information from all physical states, including the vacuum and all excited states. We are largely motivated by the interesting precedent in the literature for computing the partition function in the approximation of the free field theory partition function, imposing the singlet projection of $\mathcal{N}=4$ SYM's at high temperatures to probe the Hagedorn phase transition \cite{Sundborg:1999ue, Aharony:2003sx}. 
Unlike the SCI, we make a clear distinction between general supersymmetric (SUSY) operators and the more restricted class of BPS operators due to the temperature dependence of the partition function. We define BPS operators as a protected subsector of SUSY operators that satisfy not only the supersymmetry conditions but also an additional constraint valid at zero temperature. This relationship is illustrated in Fig.~\ref{fig:parameter space}. The partition function studied the configurations that reside on the SUSY surface but that in the high-temperature limit move away from the zero-temperature surface. This treatment is similar to some previous discussions in the literature \cite{Nian:2020qsk, David:2020ems}, where configurations similar to Fig.~\ref{fig:parameter space} have been explicitly presented.

Despite the difference, we should emphasize that the ABJM's free partition function in this paper still accounts for the same BPS operators as the SCI because we simply change the relative sign between the bosonic and the fermionic contributions of the SCI, which hinders their cancellations. Essentially, we keep single letters with $\{Q, S\}=\epsilon-h_3-j_3=0$ and proceed to count the free theory partition function from them. Generally speaking, the full partition function should trace over both BPS and non-BPS sectors in the Hilbert space. In this paper, we restrict the partition function to include only the microstates of the BPS sector, rather than counting all states. This simplification is based on the assumption that on the gravity side, the high-temperature AdS$_4$ black hole can be obtained by smoothly extrapolating the BPS AdS$_4$ black hole solution without any phase transitions. Consequently, the operators that contribute the most could be identified with those contributing to the BPS AdS$_4$ black hole.

As discussed for the case of SCI~\cite{GonzalezLezcano:2022hcf}, the Casimir energy term $Z_0$ will only contribute at subleading order in  the large-$N$ limit, since it is a function of $\mathfrak{m}_i-\mathfrak{m}_j$ and $\tilde{\mathfrak{m}}_i-\tilde{\mathfrak{m}}_j$.  
 With this foreknowledge, at leading order, we will ignore this Casimir energy term from now on.


\subsection{Large-$N$ Limit }
The manipulation of the matrix model presented above involves both the large-$N$ limit and the Cardy-like limit. A rigorous discussion of the interplay between these two limits was recently presented in~\cite{GonzalezLezcano:2022hcf}, which applies the Cardy-like expansion after taking the large-$N$ limit; other relevant discussions can be found in \cite{Hosseini:2019iad, Bobev:2024mqw}.

To connect with the supergravity description, as expected in the AdS/CFT correspondence, we consider the large-$N$ limit of the ABJM matrix model \eqref{Eq:PartitionFunction}. The standard treatment for the large-$N$ limit is to replace the Coulomb branch variables $(\alpha_i, \tilde{\alpha}_i)$ and the magnetic fluxes $(\mathfrak{m}_i, \tilde{\mathfrak{m}}_i)$ by some continuous variables~\cite{Pasquetti:2019uop}\cite{ Choi:2019zpz, GonzalezLezcano:2022hcf}
\begin{align}
    &\mathfrak{m}_i\to\frac{N^\alpha}{\beta}x_i\,,\ \alpha_i\to y_i\,;\nonumber\\
    & \tilde{\mathfrak{m}}_i\to\frac{N^\alpha}{\beta}x_i\,,\ \tilde{\alpha}_i\to \tilde{y}_i\,,
   \end{align} 
  where the value of $\alpha$ can be determined numerically from the saddle-point equations to be $\alpha=1/2$. Alternatively, $\alpha$ can be determined by the condition that all terms contribute to the same order in $N$ in the saddle-point equations, as originally discussed, for example, in~\cite{Herzog:2010hf}. 
   We also define $\delta y\equiv \tilde{y}_i-y_i$ in the following. Then, the discrete summation over $i$ and $j$ is replaced by the continuous integral using \cite{Choi:2019zpz,GonzalezLezcano:2022hcf}
   \begin{align}
    \sum_{i,j}e^{-nt\beta |\mathfrak{m}_i-\tilde{\mathfrak{m}}_j|\pm in(\tilde{\alpha}_j-\alpha_i)}\to\frac{2N^{2-\alpha}}{nt}\int dx\rho^2(x)e^{\pm in\delta y}\,,
   \end{align}
   where $t$ is an arbitrary numerical constant.
   After the replacement, the 1-loop part of the partition function can be written as
   \begin{align}
    &\log Z_{\text{ch}}=2N^{3/2}\int dx\rho^2(x)\sum_{n=1}^\infty\Biggl[\biggl(\frac{t_{3,4}^nq^{3n/2}e^{in\delta y}}{n^2(q^n-q'^n)(1-q^{2n})}+\frac{(-t_{3,4})^nq^{3n/2}q'^ne^{in\delta y}}{n^2(q^n-q'^n)(1-q^{2n})}\nonumber\\
    &\qquad\qquad\qquad\qquad\qquad\qquad\quad+\frac{t_{1,2}^nq^{3n/2}q'^ne^{in\delta y}}{n^2(q^n-q'^n)(1-q^{2n})}+\frac{(-t_{1,2})^nq^{5n/2}e^{in\delta y}}{n^2(q^n-q'^n)(1-q^{2n})}\nonumber\\
    &\qquad\qquad\qquad\qquad\qquad\qquad\quad+\frac{t_{1,2}^{-n}q^{3n/2}e^{-in\delta y}}{n^2(q^n-q'^n)(1-q^{2n})}+\frac{(-t_{1,2})^{-n}q^{3n/2}q'^ne^{-in\delta y}}{n^2(q^n-q'^n)(1-q^{2n})}\nonumber\\
    &\qquad\qquad\qquad\qquad\qquad\qquad\quad+\frac{t_{3,4}^{-n}q^{3n/2}q'^ne^{-in\delta y}}{n^2(q^n-q'^n)(1-q^{2n})}+\frac{(-t_{3,4})^{-n}q^{5n/2}e^{-in\delta y}}{n^2(q^n-q'^n)(1-q^{2n})}\biggr)\nonumber\\
    &\qquad\qquad-\Bigl(1+(-1)^n\Bigr)\frac{1+\lambda}{1-\lambda}\biggl(\frac{t_{3,4}^nq'^nq^{n/2}e^{in\delta y}}{n^2(q^n-q'^n)(1-q'^{2n})}+\frac{t_{1,2}^nq'^{2n}q^{n/2}e^{in\delta y}}{n^2(q^n-q'^n)(1-q'^{2n})}\nonumber\\
    &\qquad\qquad\qquad\qquad\qquad\qquad\quad+\frac{t_{1,2}^{-n}q'^nq^{n/2}e^{-in\delta y}}{n^2(q^n-q'^n)(1-q'^{2n})}+\frac{t_{3,4}^{-n}q'^{2n}q^{n/2}e^{-in\delta y}}{n^2(q^n-q'^n)(1-q'^{2n})}\biggr)\Biggr]\,,
   \end{align}

\begin{align}
  &\log Z_\text{g}=4N^{3/2}\int dx\rho^2(x)\sum_{n=1}^\infty\frac{1+(-1)^n}{n^2}\biggl[\frac{q^{3n}+q^{2n}q'^n}{(q^n-q'^n)(1-q^{2n})}-\frac{1+\lambda}{1-\lambda}\Bigl(\frac{q^nq'^{2n}+q'^{3n}}{(q^n-q'^n)(1-q'^{2n})}\Bigr)\biggr]\,.
\end{align}

\subsection{Cardy-Like Limit}

To study the high-temperature behaviors of the field theory, we consider the Cardy-like limit, motivated by~\cite{DiPietro:2014bca, DiPietro:2016ond, Cassani:2021fyv} (see also~\cite{Choi:2018hmj, Honda:2019cio, ArabiArdehali:2019tdm, Kim:2019yrz, Goldstein:2019gpz, GonzalezLezcano:2020yeb, ArabiArdehali:2021nsx}),
\begin{align}\label{eq:Cardy-like limit}
   \bar{\omega} = \lambda\bar{\beta} \ll \lambda_i\, ,
  \end{align}
where $\lambda$ is a dimensionless proportionality constant of order $\mathcal{O} (1)$, and $\lambda_i$ represents other chemical potentials. In this limit, both $\beta$ and $\beta'$ can be expressed in terms of $\bar{\beta}$ 
  \begin{align}
    &\beta=\frac{1}{2}(\bar{\beta}+\bar{\omega})=\frac{1}{2}\bigl(1+\lambda\bigr)\bar{\beta}\,,\nonumber\\
    &\beta'=\frac{1}{2}(\bar{\beta}-\bar{\omega})=\frac{1}{2}(1-\lambda)\bar{\beta}\,.
  \end{align}
  Applying the Cardy-like limit, the rational fractions can be expanded as
  \begin{align}
    &\frac{q^{l_1n}q'^{l_2n}}{(q^n-q'^n)(1-q^{2n})}=-\left(\frac{1}{n^2\lambda(1+\lambda)\bar{\beta}^2}+\frac{(2+\lambda)-l_1(1+\lambda)-l_2(1-\lambda)}{2n\lambda(1+\lambda)\bar{\beta}}\right)+\mathcal{O}(\bar{\beta}^0)\,,\\
    &\frac{q^{l_1n}q'^{l_2n}}{(q^n-q'^n)(1-q'^{2n})}=-\left(\frac{1}{n^2\lambda(1-\lambda)\bar{\beta}^2}+\frac{(2-\lambda)-l_1(1+\lambda)-l_2(1-\lambda)}{2n\lambda(1-\lambda)\bar{\beta}}\right)+\mathcal{O}(\bar{\beta}^0)\,.
  \end{align}
  After the above manipulations, the partition function \eqref{Eq:PartitionFunction} becomes
 \begin{align}
  Z=\int \mathcal{D}\rho\, \mathcal{D}\delta y\, \exp\Bigl(-W_{\text{eff}}[x;\rho,\delta y]\Bigr)\,,\label{eq:partition function}
 \end{align}
 \vspace*{-\baselineskip}
 with the effective action 
 
 \begin{align}
  &W_{\text{eff}}[x;\rho,\delta y]=2N^{3/2}\int dx\Biggl(\frac{ik\rho (x\,\delta y-\mu)}{\bar{\beta}(1+\lambda)}\nonumber\\
  &+\rho^2(x)\biggl[\frac{1}{\lambda(1+\lambda)\bar{\beta}^2}\left(\text{Li}_4[e^{i(\lambda_i+\delta y)}]+\text{Li}_4[e^{-i(\lambda_i+\delta y)}]+\text{Li}_4[e^{i(\pi+\lambda_i+\delta y)}]+\text{Li}_4[e^{-i(\pi+\lambda_i+\delta y)}]\right)\nonumber\\
  &-\frac{1+\lambda}{8\lambda(1-\lambda)^2\bar{\beta}^2}\left(\text{Li}_4[e^{2i(\lambda_i+\delta y)}]+\text{Li}_4[e^{-2i(\lambda_i+\delta y)}]\right)+\frac{4\text{Li}_4[1]}{8\lambda(1+\lambda)\bar{\beta}^2}-\frac{4(1+\lambda)\text{Li}_4[1]}{8\lambda(1-\lambda)^2\bar{\beta}^2}\nonumber\\
  &+\frac{2+\lambda-\frac{3}{2}(1+\lambda)}{2\lambda(1+\lambda)\bar{\beta}}\Bigl(\text{Li}_3[e^{i(\lambda_{3,4}+\delta y)}]+\text{Li}_3[e^{-i(\lambda_{1,2}+\delta y)}]\Bigr)\nonumber\\
  &+\frac{2+\lambda-\frac{1}{2}(1+\lambda)-(1-\lambda)}{2\lambda(1+\lambda)\bar{\beta}}\Bigl(\text{Li}_3[e^{i(\pi+\lambda_{3,4}+\delta y)}]+\text{Li}_{3}[e^{-i(\pi+\lambda_{1,2}+\delta y)}]\Bigr)\nonumber\\
  &+\frac{2+\lambda-\frac{3}{2}(1+\lambda)-(1-\lambda)}{2\lambda(1+\lambda)\bar{\beta}}\Bigl(\text{Li}_3[e^{i(\lambda_{1,2}+\delta y)}]+\text{Li}_3[e^{-i(\lambda_{3,4}+\delta y)}]\Bigr)\nonumber\\
  &+\frac{2+\lambda-\frac{5}{2}(1+\lambda)}{2\lambda(1+\lambda)\bar{\beta}}\Bigl(\text{Li}_3[e^{i(\pi+\lambda_{1,2}+\delta y)}]+\text{Li}_3[e^{-i(\pi+\lambda_{3,4}+\delta y)}]\Bigr)\nonumber\\
  &-\frac{2-\lambda-\frac{1}{2}(1+\lambda)-(1-\lambda)}{4\cdot 2\lambda(1-\lambda)\bar{\beta}}\frac{1+\lambda}{1-\lambda}\Bigl(\text{Li}_3[e^{2i(\lambda_{3,4}+\delta y)}]+\text{Li}_3[e^{-2i(\lambda_{1,2}+\delta y)}]\Bigr)\nonumber\\
  &-\frac{2-\lambda-\frac{1}{2}(1+\lambda)-2(1-\lambda)}{4\cdot 2\lambda(1-\lambda)\bar{\beta}}\frac{1+\lambda}{1-\lambda}\Bigl(\text{Li}_3[e^{2i(\lambda_{1,2}+\delta y)}]+\text{Li}_3[e^{-2i(\lambda_{3,4}+i\delta y)}]\Bigr)\nonumber\\
  &+\Bigl(\frac{2(2+\lambda-3(1+\lambda))}{2\lambda(1+\lambda)\bar{\beta}}+\frac{2(2+\lambda-2(1+\lambda)-(1-\lambda))}{2\lambda(1+\lambda)\bar{\beta}}\Bigr)\frac{\text{Li}_3[1]}{4}\nonumber\\
  &-\Bigl(\frac{2(2+\lambda-(1+\lambda)-2(1-\lambda))}{2\lambda(1-\lambda)\bar{\beta}}\frac{1+\lambda}{1-\lambda}+\frac{2(1+\lambda-3(1-\lambda))}{2\lambda(1-\lambda)\bar{\beta}}\frac{1+\lambda}{1-\lambda}\Bigr)\frac{\text{Li}_3[1]}{4}\biggr]+\mathcal{O}(\bar{\beta}^0)\Biggr)\,.\label{eq:effective action}
 \end{align}
 Here, we have replaced $t_i$ with $\lambda_i$ by $t_i=e^{i\lambda_i},\, (i=1,\cdots, 4)$. 
   To manipulate the above expressions, we have introduced the polylogarithmic functions defined by
   \begin{align}
    \text{Li}_n[z]=\sum_{k=1}^\infty\frac{z^k}{k^n}\,,
   \end{align}
   and used the following identity of these functions
   \begin{align}
    \text{Li}_n[z] + \text{Li}_n[-z] = 2^{1-n}\, \text{Li}_n[z^2]\,.
   \end{align}
 The constant $\mu$ is introduced as a Lagrange multiplier to impose the normalization condition for the density of states $\int dx\rho(x)=1$, just as in the case for the SCI \cite{Benini:2015eyy, GonzalezLezcano:2022hcf}.
 From this expression, we can find that the leading order of the partition function as well as the corresponding entropy is $\sim \bar{\beta}^{-2}$ at high temperature. Since the integration measure is independent of $\bar{\beta}$, we can safely neglect it hereafter.

\begin{figure}[htb!]
    \begin{center}
      \includegraphics[width=0.45\textwidth]{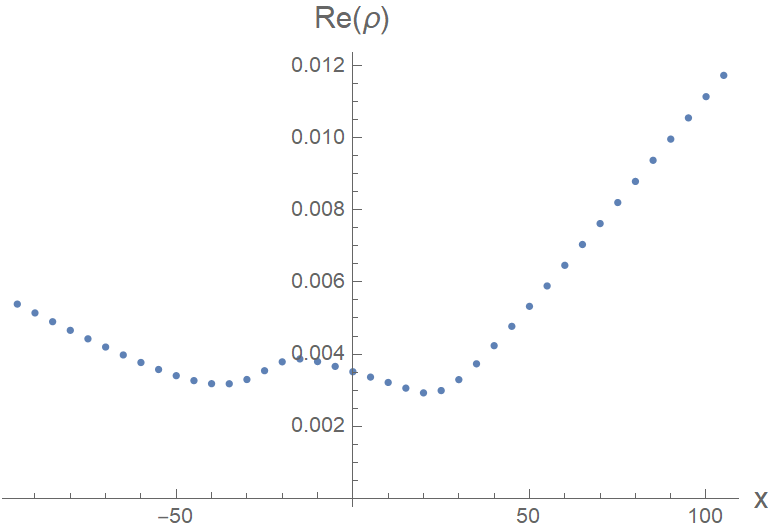}\quad
      \includegraphics[width=0.45\textwidth]{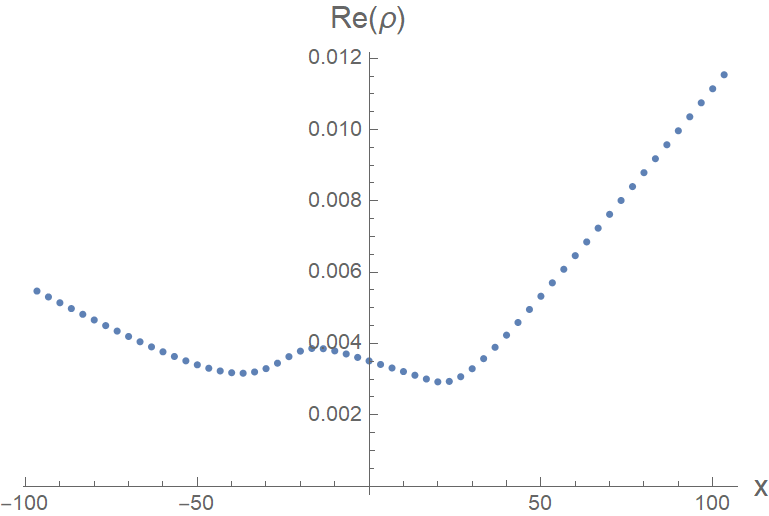}\\
    
      \includegraphics[width=0.45\textwidth]{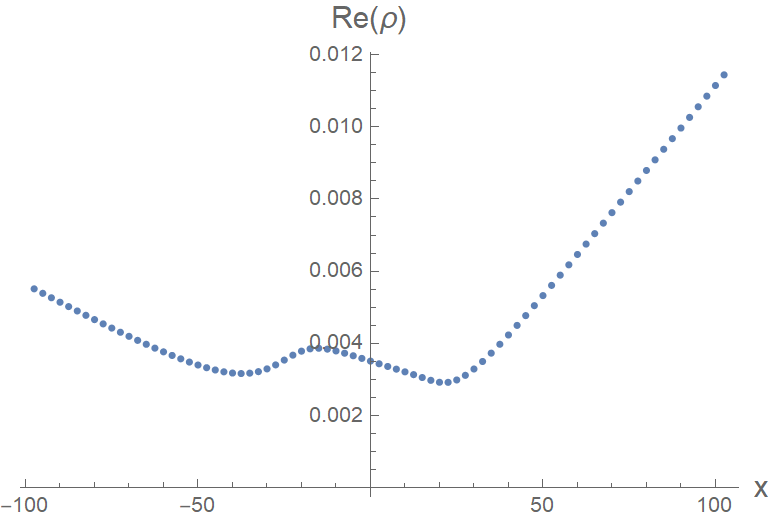}\quad
  \includegraphics[width=0.45\textwidth]{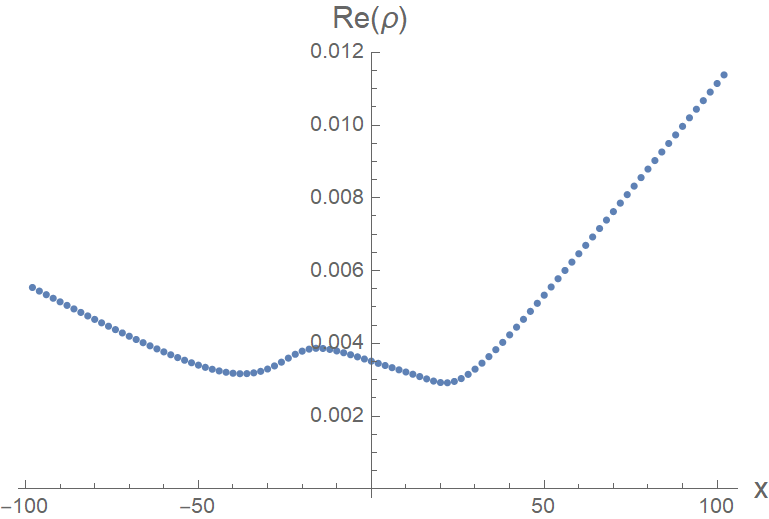}
    
    \end{center}
    \caption{\label{fig:rerho} Numerical saddle-point solution to $\text{Re}(\rho(x))$ in which the range of $x$ is divided into 40, 60, 80, 100 segments, respectively. Accordingly, the normalization condition $\int^{100}_{-100} dx\rho(x)$ is approximated by discrete summations, and the result turns out to converge to $\int^{100}_{-100} dx\rho(x)=1.01858, 1.00804, 1.00297, 1.00000$, respectively.}
\end{figure}

\begin{figure}[htb!]
    \begin{center}
      \includegraphics[width=0.45\textwidth]{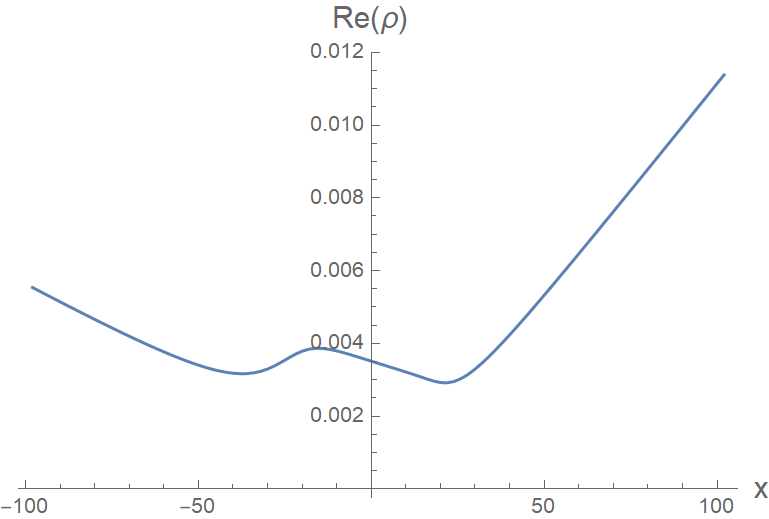}\quad
      \includegraphics[width=0.45\textwidth]{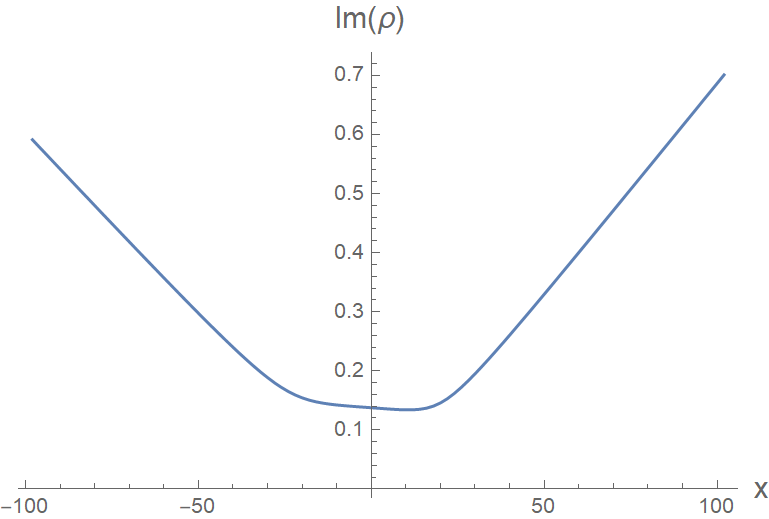}\\
    
      \includegraphics[width=0.45\textwidth]{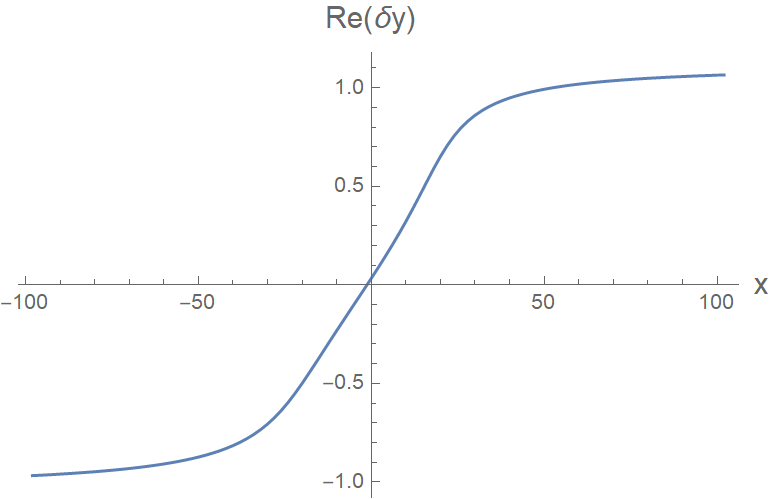}\quad
  \includegraphics[width=0.45\textwidth]{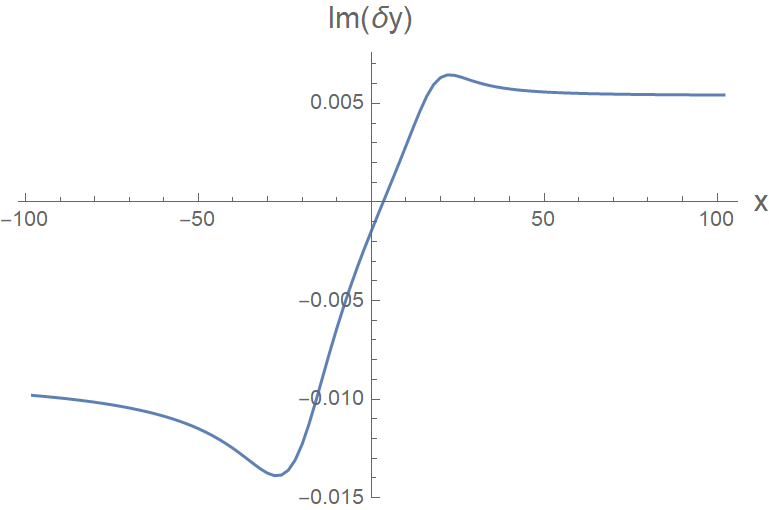}
    
    \end{center}
    \caption{\label{fig:saddle} Numerical saddle-point solutions for the matrix model of ABJM free partition function on $S^2\times S^1$. We choose $k=1$,  $\bar{\beta}=0.1$, $\lambda=0.3$, $\lambda_1=\lambda_2=\pi/2-0.2$, $\lambda_3=\lambda_4=\pi/2+0.2$. $\mu=10.7076$ is taken such that the normalization condition for the real part of the density of states $\int dx\rho(x)=1$ is satisfied.}
  \end{figure}

Varying the effective action~\eqref {eq:effective action} with respect to $\rho(x)$ and $\delta y(x)$ gives the saddle-point equations. 
To numerically solve the saddle-point equations, the interval of the variable $x$ is set to be $[-100,100]$ and evenly divided into segments, so that the integral in \eqref{eq:effective action} is approximated as a discrete summation over the segments. Given the values of $x$ for the midpoints of segments, the values of $(\rho,\delta y)$ are obtained using the FindRoot routine of Mathematica. The normalization condition for the real part of the density of states is satisfied in terms of appropriately chosen $\mu$ when given the chemical potentials $\lambda_i$ and other parameters $k$, $\bar{\beta}$, $\lambda$. We have checked the convergence of this approach by tracking the convergence of the normalization condition as more segments are added, as illustrated in Fig.~\ref{fig:rerho}. The final numerical results in Fig.~\ref{fig:saddle} for the saddle-points of the functions $\rho(x)$ and $\delta y(x)$ can be obtained by interpolating the values at a large number of discrete points.

Substituting the saddle points into Eq.~\eqref{eq:partition function}, we finally obtain the entropy in terms of chemical potentials and temperature. To compare with the entropy from the gravity~\eqref{Bekenstein-Hawking entropy2}, one can identify $\lambda_i$ ($i=1,\cdots,4$) as $\Phi_i$ of the black hole. 
   Using Eq.~\eqref{eq:phi in k from gravity side}, the entropy can be expressed as a function of $k_1$ parameterized by the temperature.

Similar to the $\mathcal{N}=4$ SYM's free partition function \cite{Sundborg:1999ue, Aharony:2003sx}, we do not expect the ABJM free partition function to capture precisely the dual AdS$_4$ black hole entropy. However, we still find some interesting qualitative agreement. Indeed, the black hole entropy from the boundary CFT's free partition function behaves as $N^{3/2}\, T_{\text{H}}^2$, which is consistent with the AdS$_4$ black hole's Bekenstein-Hawking entropy at high temperature according to the AdS$_4$/CFT$_3$ dictionary, $\frac{1}{G_N}=\frac{2\sqrt{2}}{3}g^2k^{1/2}N^{3/2}$.

\begin{figure}[htb!]
    \begin{center}
      \includegraphics[width=0.65\textwidth]{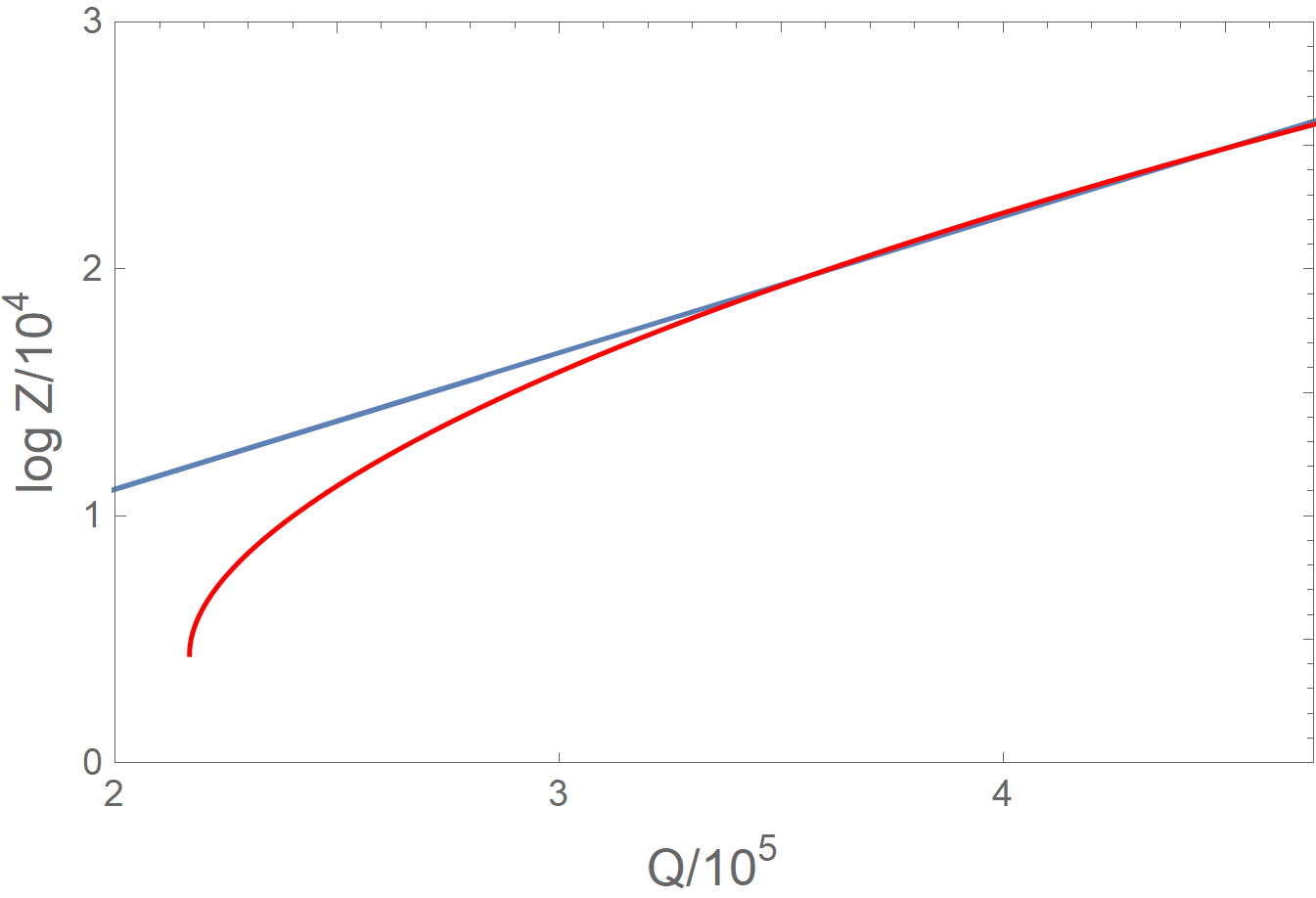}
    \end{center}
    \caption{\label{fig:double plot} $\textrm{log}\, Z$ as a function of $Q$. We choose $k=1$, $T=10$, $\lambda=0.3$ and $a=0.9999$. The range of the integration variable $x$ is fixed to be $[-100,100]$. The blue and the red curves denote the results from gravity and the ABJM free partition function on $S^2\times S^1$, respectively.}
\end{figure}

Fig.~\ref{fig:double plot} summarizes the main results of this section. It shows a qualitative agreement between the free partition function and the gravity results obtained previously. We focus on the dependence of the partition function as a function of the conserved charge $Q$, because the conserved charge is always considered to be the same quantity that connects gravity theory and field theory. The natural starting point in field theory is the grand-canonical partition function, which depends on the chemical potentials. We have to perform a Legendre transform to make it $Q_i$-dependent. According to the definition of the partition function, $Z \sim \sum e^{-\bar{\beta} E + \bar{\omega} J + \lambda_i Q_i}$, in the Cardy-like limit, the conserved charge on the field theory side can be approximated as
\begin{align}
  Q_i=\frac{\partial \log Z}{\partial \lambda_i}\,.
\end{align} 
In this way, $\textrm{log}\, Z$ as a function of chemical potential $\lambda_{1,2}$, i.e., Eq.~\eqref{eq:partition function} with Eq.~\eqref{eq:effective action}, can be changed into a function of conserved charge $Q$ conjugated with $\lambda_{1,2}$,\footnote{Note that we have imposed $\sum_{i}\lambda_i=2\pi$ and set $\lambda_1=\lambda_2$, $\lambda_3=\lambda_4$.} shown as the red curve in Fig.~\ref{fig:double plot}. On the gravity side, we impose only the gravitational SUSY condition \eqref{eq:SUSY condition}, without imposing the extremal condition \eqref{eq: zero-temperature condition}, similar to the near-extremal AdS$_4$ black hole case discussed in \cite{David:2020jhp}. In this way, we can write the partition function~\eqref{eq: grand-canonical partition function} as a function of $Q$ using Eq.~\eqref{eq: m}, \eqref{eq:TH in k from gravity side}, and \eqref{eq:Q}. The gravity result is plotted as the blue line in Fig.~\ref{fig:double plot}. As the large black hole limit is applied on the gravity side, which is equivalent to the infinite charge limit, we do not expect any results consistent with those from field theory. At large charges, we find that the gravity and the field theory results fit well up to a numerical constant,\footnote{We have shifted $\log Z$ from the matrix model by a numerical constant $2.1\times 10^4$.} supporting the legitimacy of the non-extremal AdS$_4$ black hole microstate counting via the boundary theory's free partition function.

As the charge $Q$ is increased, the quantitative agreement between the gravitational result and the matrix-model computation begins to weaken. This is expected, since our field-theoretic calculation is carried out at finite $Q$ and employs a finite integration range for the variable $x$.

\begin{figure}[htb!]
    \begin{center}
      \includegraphics[width=0.65\textwidth]{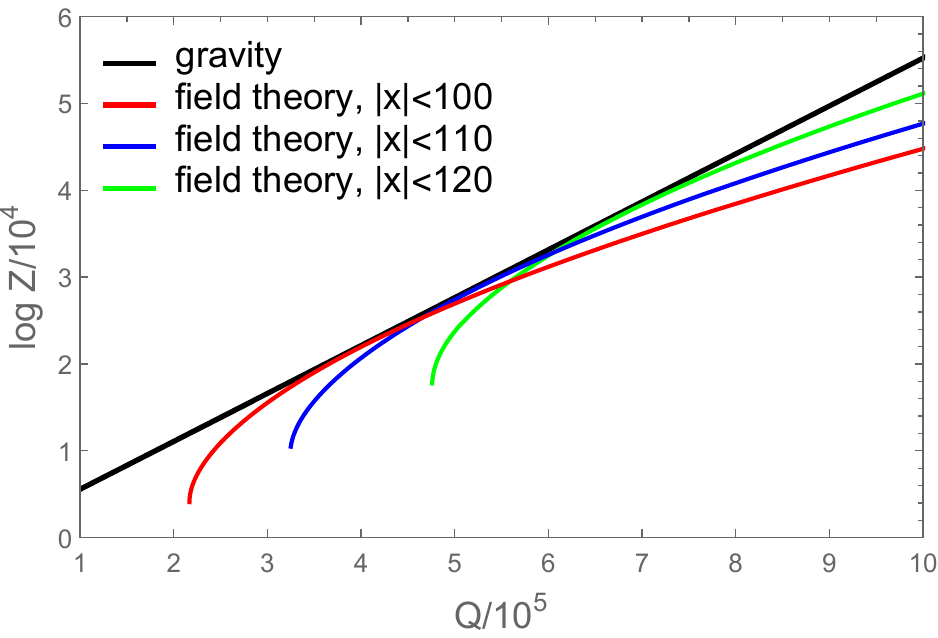}
    \end{center}
    \caption{\label{fig:double plot2}$\textrm{log}\, Z$ as a function of $Q$. We choose $k=1$, $T=10$, $\lambda=0.3$, and $a=0.9999$. The red, blue, and green curves correspond to integration ranges for the variable $x$ fixed to be $[-100,100]$, $[-110,110]$, and $[-120,120]$, respectively.}
\end{figure}

 However, when we extend the integration range of $x$ to larger intervals, the region in which the gravity and field-theory curves overlap shifts systematically toward larger values of $Q$, as illustrated in Fig.~\ref{fig:double plot2}. This consistent shift suggests that in the formal limit where the integration range is taken to infinity, the field-theory result is expected to converge and match the gravity prediction precisely in the corresponding infinite-charge limit. The deviation seen at finite $Q$ therefore reflects finite-cutoff effects in our numerical setup, rather than a fundamental disagreement. This asymptotic trend reinforces the conclusion that the universal scaling captured by the free partition function describes the correct macroscopic behavior of the dual black hole.

 In this section, we report the results of our explorations, i.e., the 3d free BPS partition function can capture the main features of AdS$_4$ supersymmetric black hole entropy at high temperatures up to some overall constant factors. A deep understanding of these results is a much harder task that we do not attempt to tackle in this paper. It is worth highlighting that there are numerous precedents in the literature where an {\it a priori} naive counting of degrees of freedom has yielded surprising results \cite{Skagerstam:1983gv, Sundborg:1999ue, Aharony:2003sx, Drukker:2010nc, Bobev:2023lkx, Fluder:2020pym}. Ultimately, the underlying ``reason'' has been formulated as the power of the singlet condition in the path integral to capture nontrivial physics.

\section{Hawking Radiation}\label{sec:Hawking radiation}

We can see from the previous sections that various approaches lead to similar results for the non-extremal AdS$_4$ black holes, namely, $S_{\text{BH}}^{(1)} = S_{\text{BH}}^{(2)} = S_{\text{BH}}^{(3)}$ as anticipated in the introduction. In particular, the covariant phase space formalism and the previous hidden conformal symmetry allow us to formulate the non-extremal AdS$_4$ black hole entropy as a CFT$_2$ entropy, which receives contributions from both the left-moving and the right-moving sectors. Based on this formulation, we can provide a microscopic foundation for the Hawking radiation rate of a non-extremal AdS$_4$ black hole. This microscopic method was initiated in \cite{Callan:1996dv} for the D1/D5 CFT and in \cite{Emparan:1998qp} for the BTZ black hole. Later, it was generalized to near-extremal AdS black holes in \cite{Nian:2020qsk, David:2020jhp} using the Kerr/CFT correspondence. In this section, we will make a further generalization to non-extremal AdS$_4$ black holes, similar to the treatments in \cite{Nian:2023dng, Nian:2023xmr}.

As shown in \cite{Callan:1996dv, Nian:2020qsk, David:2020jhp, Nian:2023xmr}, once we have a CFT$_2$ formulation of the higher-dimensional black hole entropy, we can interpret the Hawking radiation as the scattering between the left-moving and the right-moving modes in CFT$_2$. The Hawking radiation rate is just the scattering rate, which can be computed in a standard way. Following \cite{Callan:1996dv, Nian:2020qsk, David:2020jhp}, we compare the non-extremal AdS$_4$ black hole entropy in canonical and microcanonical ensembles:
\begin{align}
  S_{BH} & = \frac{\pi^2}{3} c_L T_L + \frac{\pi^2}{3} c_R T_R \\
  {} & = 2 \pi \sqrt{\frac{c_L N_L}{6}} + 2 \pi \sqrt{\frac{c_R N_R}{6}}\, ,	
\end{align}
and obtain the relations
\be\label{eq:T and N relations}
  T_L = \frac{1}{\pi} \sqrt{\frac{6 N_L}{c_L}}\, ,\quad T_R = \frac{1}{\pi} \sqrt{\frac{6 N_R}{c_R}}\, .
\ee
Using the expressions of $c_{L, R}$ \eqref{eq:cLcR} and $T_{L, R}$ \eqref{eq:TL} \eqref{eq:TR} from the covariant phase space formalism, we can obtain explicit expressions for $N_{L, R}$. In the high-temperature, non-extremal limit ($a \to 0$), there is the relation $N_L \approx N_R$, i.e., both the left-moving and the right-moving sectors are highly excited.

In the microcanonical ensemble, the partition function of the right-moving sector can be written as
\be
  Z_R = \sum_{N_R} q^{N_R}\, d(N_R) = \sum_{N_R} q^{N_R}\, e^{S_R} = \sum_{N_R} q^{N_R}\, e^{2 \pi \sqrt{c_R\, N_R / 6}}\, ,
\ee
where $q \equiv e^{-1 / T_R}$, and $d(N_R)$ denotes the degeneracy of the right-moving modes. Performing the saddle-point approximation, we obtain
\be
  N_R = q \frac{\partial}{\partial q}\, \textrm{log}\, Z_R \approx \frac{\pi^2 c_R}{6\, \left(\textrm{log} (q) \right)^2}\, .
\ee
Similar results apply to the left-moving sector:
\be
  N_L = q \frac{\partial}{\partial q}\, \textrm{log}\, Z_L \approx \frac{\pi^2 c_L}{6\, \left(\textrm{log} (q) \right)^2}\, .
\ee
These saddle points consistently reproduce the results \eqref{eq:T and N relations}.

The occupation numbers are given by the Bose-Einstein statistics:
\begin{align}
  \rho_L (k_0) & = \frac{q^n}{1 - q^n} = \frac{e^{- k_0 / T_L}}{1 - e^{- k_0 / T_L}}\, ,\\
  \rho_R (k_0) & = \frac{q^n}{1 - q^n} = \frac{e^{- k_0 / T_R}}{1 - e^{- k_0 / T_R}}\, ,
\end{align}
where $n$ denotes the momentum quantum number of the modes moving in the time circle of the AdS$_4$ black hole near-horizon region, while $k_0$ stands for the typical energy of these modes. Here, we assume $n = k_0$ for simplicity in the above expressions. At high temperatures, $k_0 \ll T_L \approx T_R$, and the occupation numbers can be approximated as
\be\label{eq:rho high temperature}
  \rho_L (k_0) = \frac{e^{- k_0 / T_L}}{1 - e^{- k_0 / T_L}} \approx \frac{T_L}{k_0}\, ,\quad \rho_R (k_0) = \frac{e^{- k_0 / T_R}}{1 - e^{- k_0 / T_R}} \approx \frac{T_R}{k_0}\, .
\ee
Alternatively, they are also both approximately the Bose-Einstein distribution with the physical Hawking temperature $T_H$:
\be\label{eq:rho distribution factor}
  \rho_L (k_0) \approx \rho_R (k_0) \approx \frac{e^{- 2 k_0 / T_H}}{1 - e^{- 2 k_0 / T_H}}\, ,
\ee
where the physical momentum $2 k_0$ receives contributions from both the left-moving and the right-moving sectors.

Finally, as discussed in \cite{Callan:1996dv, Nian:2020qsk, David:2020jhp}, the Hawking radiation rate for non-extremal AdS$_4$ black holes can be obtained as
\be\label{eq:RadiationRate}
  d\Gamma \sim \frac{d^4 k}{2 k_0} \frac{1}{p_0^L\, p_0^R}\, |\mathcal{A}|^2\, c_L\, c_R\, \rho_L (k_0)\, \rho_R (k_0)\, ,
\ee
where $c_L c_R$ is approximately the total degrees of freedom at a given quantum number $n$, while $\mathcal{A}$ stands for the scattering amplitude of the left-moving and the right-moving modes in the near-horizon CFT$_2$. In the expression \eqref{eq:RadiationRate}, we can keep either $\rho_L$ or $\rho_R$ for the statistical distribution factor \eqref{eq:rho distribution factor}, while approximating the other as \eqref{eq:rho high temperature} in the high-temperature limit. Either choice leads to the universal form of the Hawking radiation rate as
\be
  d\Gamma \sim (\text{horizon area})\cdot \frac{e^{-2 k_0 / T_H}}{1 - e^{- 2 k_0 / T_H}}\, d^4 k\, ,
\ee
where we have applied the Cardy formula $S_{BH} = \frac{\pi^2}{3} c_L T_L = \frac{\pi^2}{3} c_R T_R \sim $ (horizon area). Similar expressions have been previously found in the literature for various near-extremal black holes \cite{Callan:1996dv, Nian:2020qsk, David:2020jhp}. Hence, the low-energy Hawking radiation rate can be computed from a near-horizon CFT$_2$ approach, revealing again the hidden conformal symmetry of the non-extremal AdS$_4$ Kerr-Newman black holes.

Although our primary goal has been to explore far-from-extremality black holes, it would be interesting to connect this CFT$_2$ approach to Hawking radiation with recent approaches to modifications of Hawking radiation rooted in the effective quantum Jackiw-Teitelboim description of the throat region for near-extremal black holes \cite{Brown:2024ajk} with emphasis on the rotating black holes \cite{Maulik:2025hax}.

\section{Discussion}\label{sec:discussion}

In this manuscript, we have generalized various approaches to the Bekenstein-Hawking entropy of electrically charged rotating AdS$_4$ black holes from the near-extremal case to the far-from-extremality regime by using macroscopic and microscopic frameworks, including: (i) the covariant phase space formalism and (ii) the free partition function of the boundary ABJM theory. The agreement between these two approaches supports certain universality in the entropy of AdS black holes. We also probed the universality of the entropy of AdS$_4$ black holes at high temperatures by explicitly considering the thermodynamics from a boundary fluid dynamics point of view. Finally, we discussed the Hawking radiation rate from a near-horizon CFT$_2$ picture. Each point of view contributed to a coherent picture for far-from-extremality black holes. \\

We first considered the near-horizon region where the geometry of the rotating black hole is warped AdS$_3$, which is dual to a two-dimensional conformal field theory. Based on the Kerr/CFT correspondence, the left- and right-moving central charges in the Virasoro algebra of the two-dimensional conformal field theory are expressed in terms of linearized charges, including the contributions from the Iyer-Wald charge and the Wald-Zoupas charge, using the covariant phase space formalism. Together with the temperatures of both the left- and right-moving sectors, we reproduced the black hole entropy using the Cardy formula in the two-dimensional conformal field theory. The result matches precisely with the Bekenstein-Hawking entropy even far away from extremality.

Going beyond the near-horizon region and focusing on the asymptotically AdS boundary, we evaluated the black hole entropy in terms of an approximation to the free partition function from the dual boundary field theory based on the AdS/CFT correspondence. According to the AdS/CFT correspondence, the partition function of the AdS$_4$ black hole we are considering is equivalent to the partition function of three-dimensional ABJM with  $\mathcal{N}=6$ supersymmetry and gauge group $U(N)_k\times U(N)_{-k}$ on $S^1\times S^2$ with appropriate boundary conditions on the fields. We approximate the partition function by essentially considering the states that contribute to the superconformal index without imposing supersymmetry, thereby removing the $(-1)^F$ factor in the trace over the Hilbert space of the theory. As a result, the partition function depends on one extra chemical potential due to the absence of boson-fermion cancellations. We use the Cardy-like limit to simplify the resulting matrix model for the free partition function. Expanding in powers of inverse temperature $\bar{\beta}$, the one-loop contributions for the partition function, including the chiral multiplets and the vector multiplets, are expressed in terms of the polylogarithmic functions. We evaluate the partition function in the large-$N$ limit by redefining the variables and varying the partition function with respect to these variables to obtain the saddle-point equations. These equations are solved numerically, and we show that the dependences on the rank of the gauge group and the temperature are consistent with the dual black hole entropy. Additionally, the fluid dynamics on the boundary, based on AdS/CFT, can also be used to describe the black hole entropy.

 The conceptual foundation of our work rests on the following points, which together justify the use for employing a free BPS partition function to explore the  black hole entropy far from extremality.

\begin{itemize}

\item Our computation directly extends the cases that have been rigorously validated for extremal~\cite{David:2020ems}  and near-extremal~\cite{Nian:2020qsk, David:2020jhp}  BPS black holes.  Our approach adopts the same underlying spectrum of BPS-derived degrees of freedom, but relaxes the $(-1)^F$ grading to allow their thermal contribution at finite temperature, thereby generalizing a proven framework into the high-temperature domain.

\item In the high-temperature regime, the leading behavior of thermodynamic quantities is governed by universal data, i.e., the number of degrees of freedom and the symmetries of the system, rather than by interaction-dependent details.

\item Our analysis assumes that no reorganizing phase transition, such as a Hagedorn transition or a black-hole/string transition, occurs between the BPS extremal limit and the high-temperature black-hole regime we study. Thus, the microscopic constituents do not undergo a drastic reorganization between the free theory and the strongly coupled regime.

\end{itemize}

A description of the black hole in terms of a CFT$_2$ associated with the near-horizon region provides a natural path to the computation of Hawking radiation. We exploited this path and obtained a fairly universal expression for the Hawking radiation rate, which is proportional to the area of the event horizon.

There are several interesting open problems that we have postponed for future studies. First, it would be interesting to achieve a more precise evaluation of the large-$N$ field-theoretic partition function. One path would be to develop a systematic technique to more efficiently solve the saddle-point equations, making full use of the properties of the polylogarithm functions \cite{Zagier:2007knq}. Second, it would be interesting to perform a finite-$N$ counterpart of the analysis; this would go beyond the saddle-point approximation and possibly provide input into the structure of sub-leading terms in the entropy. There is, of course, the numerical obstacle of dealing with the sum over the dynamical magnetic fluxes. In addition, there are various open problems of a more conceptual nature. It would be interesting to reproduce subleading corrections to the black hole entropy from different microscopic approaches, which address the nature of the underlying degrees of freedom responsible for the entropy more directly. It is also natural to generalize these frameworks to study the microscopic counting of black holes in other dimensions, where near-extremal descriptions are already available. In addition to aspects of black hole entropy, a more interesting question is to focus on the dual field theory descriptions afforded by the AdS/CFT correspondence and characterize the corresponding renormalization group flow between CFT$_d$ at the boundary of AdS$_{d+1}$ spacetime and CFT$_2$ near the horizon of the asymptotically rotating AdS black hole.

\acknowledgments

We would like to thank Evan Deddo, Sameer Murthy, Tomoki Nosaka, Cheng Peng, Alessia Segati, and Minwoo Suh for many helpful discussions.
 We thank an anonymous referee for valuable comments and suggestions. This work is supported in part by the NSFC under grants No.~12375067, No.~12147103, and No.~12247103. LAPZ is partially supported by the U.S. Department of Energy under grant DE-SC0007859.  All three authors would like to thank PCFT in Hefei for the warm hospitality during the final stage of this work.



\bibliographystyle{utphys}
\bibliography{ref}



\end{document}